\documentclass[12pt]{article}
\usepackage[centertags]{amsmath}
\usepackage{amsthm}
\usepackage{newlfont}
\usepackage{amsmath}
\usepackage{amsfonts,amsmath,latexsym}
\usepackage{mathrsfs}
\usepackage{amssymb}
\usepackage{amsbsy}
\usepackage{indentfirst}
\usepackage{mathrsfs}
\usepackage{graphicx}
\usepackage{amssymb, amsmath}
\usepackage{bm}
\usepackage[flushleft]{threeparttable}

\newcommand{\norm}[1]{\left\lVert#1\right\rVert}

\usepackage{diagbox}
\usepackage{rotating}
\usepackage{booktabs}
\usepackage{bigstrut, multirow}
\allowdisplaybreaks[4]

\titlepage
\newlength{\defbaselineskip}
\newcommand{\be}{\begin{eqnarray}}
\newcommand{\ee}{\end{eqnarray}}
\newcommand{\bestar}{\begin{eqnarray*}}
\newcommand{\eestar}{\end{eqnarray*}}
\newcommand{\ignore}[1]{}
{} \theoremstyle{plain}
\newtheorem{thm}{THEOREM}[section]

\newtheorem{lem}{LEMMA}[section]

\theoremstyle{definition}
\newtheorem{defn}{DEFINIOTION}[section]
\newtheorem{assum}{ASSUMPTION}[section]
\newtheorem{exam}{EXAMPLE}[section]
\theoremstyle{remark}

\numberwithin{equation}{section}

\def\>{\geq}
\def\<{\leq}

\begin{document}

\baselineskip=0.7 true cm

\title{Testing error distribution by kernelized Stein discrepancy in multivariate time series models}

\author{Donghang Luo$^{1}$, Ke Zhu$^{1}$, Huan Gong$^{2}$ and Dong Li$^{2}$\footnote{Address correspondence to Dong Li:
Center for Statistical Science, Tsinghua University, Beijing, China.  E-mail: malidong@tsinghua.edu.cn
} \\
\\
\emph{University of Hong Kong$^{1}$ and Tsinghua University$^{2}$}
}
\maketitle


\begin{abstract}
Knowing the error distribution is important in many multivariate time series applications.
To alleviate the risk of error distribution mis-specification, testing methodologies are needed
to detect whether the chosen error distribution is correct. However,
the majority of the existing tests only deal with the multivariate normal distribution for some
special multivariate time series models,
and they thus can not be used to testing for the often observed heavy-tailed and skewed error distributions in applications.
In this paper, we construct a new consistent test for general multivariate time series models, based on the kernelized Stein discrepancy.
To account for the estimation uncertainty and unobserved initial values, a bootstrap method is provided to
calculate the critical values.
Our new test is easy-to-implement for a large scope of multivariate error distributions,
and its importance is illustrated by simulated and real data.
\end{abstract}

\noindent

\noindent{\it Keywords and phrases}: Consistent test; Kernelized Stein discrepancy; Multivariate time series model; Testing multivariate error distribution.

\newpage

\section{Introduction}
Consider a multivariate stationary time series $\{Y_t\}$ with $Y_t=(Y_{1t},...,Y_{dt})^\top\in\mathbb{R}^d$, and $Y_t$ admits the following specification
\begin{equation}
    \label{E1}
    Y_t=M(I_{t-1};\theta_{0})+C^{1/2}(I_{t-1};\theta_0)\eta_t,
\end{equation}
where $I_t = \{Y_t, Y_{t-1}, ...\}$ is the information set up to time $t$,
$\theta_0\in \mathbb{R}^p $
is the true yet unknown model parameter, $\eta_t \in \mathbb{R}^d $ is a sequence of independent and identically distributed (i.i.d.) errors with zero mean
and identity covariance matrix $\mathrm{I}_d$, $M(\cdot;\theta_{0})\in \mathbb{R}^d$ is a known measurable vector function indexed by $\theta_{0}$,
and $C(\cdot;\theta_0)\in\mathbb{R}^{d\times d}$ is a known measurable symmetric and positive definite matrix function indexed by $\theta_{0}$.
Let $\mathcal{F}_{t} := \sigma(I_t)$ be a sigma-field generated by $I_{t}$. Conditional on $\mathcal{F}_{t-1}$,
$M(I_{t-1};\theta_{0})$ and $C(I_{t-1};\theta_0)$ in (\ref{E1}) are the conditional mean vector and conditional covariance matrix of $Y_t$, respectively.
The general specification in (\ref{E1}) covers many often used multivariate models including, for example, the vector autoregressive and moving-average (VARMA) model, the multivariate generalized autoregressive conditional heteroskedasticity (MGARCH) model, and their variants and combinations. For surveys on the multivariate time series models, we refer to L\"{u}tkepohl (2005), Bauwens et al. (2006),
Tsay (2013), and Francq and Zako\"{i}an (2019).

For model (\ref{E1}), $\eta_t$ is assumed to have certain continuous probability density function (p.d.f.) $p_0(x)$ in a myriad of applications, which include the validity of
capital asset pricing model (Berk, 1997),
the optimal forecasts (Christoffersen and Diebold, 1997), the density forecasts (Diebold et al., 1998), the interval forecasts (Zhu and Li, 2015), the option pricing (Zhu and Ling, 2015),
and the Value-at-Risk and Expected Shortfall calculations (Taylor, 2019).
However, the true p.d.f. of $\eta_t$, denoted by $p(x)$, is generally unknown in practice, and
the empirical researchers could make wrong conclusions
if their assumed p.d.f. $p_0$ is different from $p$. Motivated by this, it is important to testing for the following hypotheses
\begin{equation}\label{Hypothesis}
    H_0: p = p_0 \text{ versus }
    H_1: p \neq p_0.
\end{equation}
Our considered hypotheses in (\ref{Hypothesis}) are designed for the unobserved model error $\eta_t$,
which nests the observed data (i.e., $\eta_t=Y_t$) as a special case. In this paper,
we mainly focus on the testing for unobserved $\eta_t$, and
the testing methodologies for univariate/multivariate observed time series can be found in
Lobato and Velasco (2004), Bai and Ng (2005), Mecklin and Mundfrom (2004), Sz\'{e}kely and Rizzo (2005),  and the references therein.


Since $\eta_t$ is unobserved, one need use the model residual $\widehat{\eta}_t$ to form valid tests for the hypotheses in (\ref{Hypothesis}).
When $Y_t$ is univariate (i.e., $d=1$), a number of different testing methods were proposed in the literature.
Bontemps and Meddahi (2005) considered the robust moment tests for normality of $\eta_t$ by using the Hermite polynomials, and their idea was further extended in Bontemps and Meddahi (2012) to examine the general distribution of $\eta_t$.
Although these robust moment tests are easy-to-implement with a chi-square limiting null distribution,
they are inconsistent as only a finite number of moments of $\eta_t$ are considered for the testing purpose.
To construct consistent tests, other strategies have been adopted.
For the general model as in (\ref{E1}), Bai (2003) developed
Kolmogorov--Smirnov (KS) and Cram\'{e}r--von Mises (CvM) tests by measuring the distance between the empirical distribution of $\widehat{\eta}_t$ and the cumulative distribution of $\eta_t$. For the GARCH model, Horv\'{a}th and Zitikis (2006)
gave a smooth-type test by measuring the distance between the kernel density estimator of $p$ and the assumed density $p_0$ in $L_{\nu}$-norm with $1<\nu<\infty$, and
Klar et al. (2012) constructed an integrated test by measuring the distance between
the empirical characteristic function of $\widehat{\eta}_t$ and the characteristic function of $p_0$.
For the ARMA--GARCH model, Koul and Ling (2006) studied a weighted KS test based on a vector of certain weighted residual empirical processes.


When $Y_t$ is multivariate (i.e., $d>1$) and both $M(\cdot;\theta_{0})$
and $C(\cdot;\theta_0)$ are constants,
most of earlier efforts were made to detect the normality of $\eta_t$. See, for example, Mardia (1974),
Henze and Zirkler (1990),  Doornik and Hansen (2008), and references therein.
When $Y_t$ is multivariate but either $M(\cdot;\theta_{0})$
or $C(\cdot;\theta_0)$ is non-constant, only few testing methods were provided for the MGARCH model.
For instance, Bai and Chen (2008) applied a similar idea as Bai (2003) to propose consistent KS tests for detecting the multivariate normal and
$t_{\nu}$ distributions of $\eta_t$. Their tests are asymptotically distribution-free, however, they are not fully consistent and require the explicit form of the conditional cumulative distribution function of $Y_{it}$ (conditional on $(Y_{1t},...,Y_{i-1,t})$), which is neither available for other multivariate distributions, nor easily computable for the dimension $d>2$. Francq et al. (2017) developed KS and CvM tests to examine whether $\eta_t$ has the elliptic distribution by extending the idea of
Henze et al. (2014).
These KS and CvM tests are not asymptotically distribution-free, and their application scope could be narrowed down
when detecting the exact distribution of $\eta_t$ is needed.
Henze et al. (2019) constructed consistent tests for the normality of $\eta_t$ by using the identity
\begin{equation}\label{Henze}
    R_{\eta}(v)M_{\eta}(v)-1=0 \mbox{ for each }v\in \mathbb{R}^d,
\end{equation}
where  $R_{\eta}(v)$ is the real part of $\varphi_{\eta}(v)$, $\varphi_{\eta}(v)$ is the characteristic function of $\eta_t$,
and $M_{\eta}(v)$ is the moment generating function of $\eta_t$. Since the identity above only holds for multivariate normal
distributions, their idea can not be extended to testing for other distributions.
In economic and financial applications, many heavy-tailed or skewed error distributions could outperform
the multivariate normal or $t_{\nu}$ distribution (see, e.g., Haas et al. (2004), Bauwens and Laurent (2005),  De Luca et al. (2006), and references therein).
Hence, it is necessary to construct a valid test
for detecting the general multivariate distribution of $\eta_t$ in model (\ref{E1}).

This paper is motivated to propose a new consistent test for $H_0$ based on the kernelized Stein discrepancy (KSD) in Liu et al. (2016).
The KSD measures the distance between the (Stein) score functions of $p$ and $p_0$ under the norm induced by a kernel function.
For the observed data $\eta_t$, Liu et al. (2016) constructed a test statistic for $H_0$, and established its asymptotics. However,
when $\eta_t$ is replaced by $\widehat{\eta}_t$, we find that their results are not applicable any more due to the estimation effect in $\widehat{\eta}_t$. To handle the estimation effect, our new KSD-based test
is constructed based on a subsample of $\widehat{\eta}_t$.
Under certain conditions, we show that our test has no estimation effect, and establish
its asymptotics under $H_0$ and $H_1$.
Although the estimation effect is negligible in theory,
it may still exist in finite samples especially when
 the sample size is small.
To overcome this difficulty, we introduce a simple parametric bootstrap method to calculate the critical values of our test.
Simulations show that our test performs well in the examined cases, even when no or few effective data $\{\widehat{\eta}_t\}$ are discarded by our
subsampling technique. A real data analysis is further given to demonstrate the usefulness of our test.

The remaining paper is organized as follows. Section 2 introduces the KSD-based test statistic.
Section 3 studies the asymptotics of the KSD-based test statistic and provides
a parametric bootstrap method to calculate the critical values. Simulation results are reported in Section 4, and
a real example is offered in Section 5. Concluding remarks are given in Section 6. Proofs are deferred into Appendices.

\section{KSD-based test statistic}

\subsection{Preliminaries on the KSD}

In this paper, we construct a new  test for hypotheses in (\ref{Hypothesis}) based on the kernelized Stein discrepancy (KSD) in Liu et al. (2016).
Let $p(x)$ be the true p.d.f. of $\eta_t$ in (\ref{E1}) with the support $\aleph\subseteq \mathbb{R}^d$.
To introduce the KSD, we first need define the (Stein) score function of $p$ and the Stein class of $p$.

\begin{defn}
The (Stein) score function of $p$ is defined as
$$s_{p}(x)=\nabla_{x}\log p(x)=\frac{\nabla_{x}p(x)}{p(x)}.$$
\end{defn}

\begin{defn}
A function $f(x):\aleph\to\mathbb{R}$ is in the Stein class of $p$
if $f$ is continuous differential and satisfies
\be \label{Stein_class}
\int_{x\in\aleph}\nabla_{x}\big(f(x)p(x)\big)dx=0.
\ee
\end{defn}
When $\aleph=\mathbb{R}^d$, by using
integration by parts, the condition (\ref{Stein_class}) holds if $$\lim_{\|x\|\to\infty}f(x)p(x)=0,$$
which holds, for example, if $p(x)$ is bounded and $\lim_{\|x\|\to\infty}f(x)=0$.

Next, let $k(x,x')$ be an integrally strictly positive definite kernel function, that is,
$$\int_{x\in\aleph}\int_{x'\in\aleph}g(x)k(x,x')g(x')dxdx'>0$$
for any function $g(x)$ satisfying $0<\|g\|_{2}^{2}<\infty$. With the kernel function $k$, we are ready to give the definition of
KSD between the distributions of $p$ and $p_0$.

\begin{defn}
The KSD $\mathbb{S}(p,p_0)$ is defined as
\be\label{KSD_1}
\mathbb{S}(p,p_0)=E_{\eta,\eta'\sim p}\big[\delta_{p_0,p}(\eta)^{\top}k(\eta,\eta')\delta_{p_0,p}(\eta')\big],
\ee
where $\delta_{p_0,p}(x)=s_{p_0}(x)-s_{p}(x)$ is the score difference between
$p_0$ and $p$, and $\eta$, $\eta'$ are i.i.d. from $p$.
\end{defn}

Clearly, the KSD $\mathbb{S}(p,p_0)$ measures the difference between the (Stein) score functions of $p$ and $p_0$ under a norm induced by the kernel function $k$. If $p$ and $p_0$ are continuous with $\|p\delta_{p_0,p}\|_{2}^{2}<\infty$,
Liu et al. (2016) showed that
$\mathbb{S}(p,p_0)\geq 0$ and
\be\label{Equiv}
\mathbb{S}(p,p_0)= 0 \mbox{ if and only if } p=p_0.
\ee
In view of the result (\ref{Equiv}), we can detect the null hypothesis $H_0$ in (\ref{Hypothesis}) by examining whether $\mathbb{S}(p,p_0)$ is significantly different from zero.
However, a direct testing implementation based on (\ref{KSD_1}) is infeasible, since the score difference $\delta_{p_0,p}$ is unknown. To overcome this difficulty, we need an additional condition on the kernel function $k$.

\begin{defn}
The kernel function $k(x,x')$ is in the Stein
class of $p$ if $k(x,x')$ has continuous second order partial derivatives, and both $k(x,\cdot)$ and $k(\cdot,x)$ are in the Stein
class of $p$ for any fixed $x$.
\end{defn}

When the kernel function $k$ is in the Stein
class of $p$, Liu et al. (2016) found that the KSD in (\ref{KSD_1}) becomes
\be\label{KSD_2}
\mathbb{S}(p, p_0)=E_{\eta,\eta' \sim p}\big[u(\eta,\eta')\big],
\ee
where
\bestar
&&u(x,x') = s_{p_0}(x)^\top k(x,x') s_{p_0}(x') + s_{p_0}(x)^\top \nabla_{x'}k(x,x') + \nabla_{x}k(x,x')^\top s_{p_0}(x') \\
&&\quad\quad\quad\quad\quad+ trace(\nabla_{x,x'}k(x,x')).
\eestar
Now, the formula of $\mathbb{S}(p, p_0)$ in (\ref{KSD_2}) is tractable for the testing purpose, since it only depends on the score function $s_{p_0}$ and the kernel function $k$, both of which are known under $H_0$.

\subsection{The KSD-based test statistic}

To form our test statistic, a sample counterpart of $\mathbb{S}(p, p_0)$ in (\ref{KSD_2}), based on the model residuals, is needed.
Let $\theta = (\theta_1, ..., \theta_p)^\top \in \Theta \subset \mathbb{R}^p $ be the unknown parameter of model (\ref{E1}), where $\Theta$ is compact parametric space. Assume that $\theta_0$ is an interior point of $\Theta$, and denote
\begin{equation} \label{E2}
    g(Y_t, I_{t-1};\theta)=C^{-1/2}(I_{t-1};\theta)\big(Y_t-M(I_{t-1};\theta)\big).
\end{equation}
By (\ref{E2}), the model residual in (\ref{E1}) can be computed as
\begin{equation}
    \label{E3}
    \widehat{\eta}_t = g(Y_t, \widehat{I}_{t-1}; \widehat{\theta}_n),
\end{equation}
where $\widehat{I}_{t}$, containing possible given initial values, is the truncated information set at time $t$, and $\widehat{\theta}_n$ is an estimator of $\theta_0$.
With model residuals $\{\widehat{\eta}_{t}\}_{t=1}^{n}$, the KSD-based test statistic $\widehat{\mathbb{S}}$ as the estimator of
$\mathbb{S}(p, p_0)$ in (\ref{KSD_2}) is given by
\begin{equation}\label{E4}
    \widehat{\mathbb{S}}= \frac{2}{n_{0}(n_{0}-1)} \sum_{n-n_0+1 \leq i < j \leq n} u(\widehat{\eta}_i, \widehat{\eta}_j),
\end{equation}
where $n_0=[K_0n^{1-\varepsilon}]$ for some $K_0, \varepsilon>0$. Clearly, $\widehat{\mathbb{S}}$ is a U-statistic with kernel function $u(\widehat{\eta}_i, \widehat{\eta}_j)$, and the calculation of $\widehat{\mathbb{S}}$ only requires the computation of $s_{p_0}$, $\nabla_{x}k(x,x')$, $\nabla_{x'}k(x,x')$, and $\nabla_{x,x'}k(x,x')$, which does not raise any computational burden even for a large dimension $d$.
For the kernel function $k$, the often used one is the Gaussian kernel
\begin{flalign}\label{Gaussian_k}
k(x,x')=\exp\Big(-\frac{1}{2\sigma^2}\|x-x'\|^{2}\Big),
\end{flalign}
where $\sigma>0$ is a fixed constant; in this case, we have
\begin{flalign*}
\nabla_{x}k(x,x')&=k(x,x')\frac{x'-x}{\sigma^2}, \,\,\, \nabla_{x'}k(x,x')=k(x,x')\frac{x-x'}{\sigma^2},\\
\nabla_{x,x'}k(x,x')&= \frac{k(x,x')}{\sigma^2}\left( \mathrm{I}_d - \frac{(x-x')(x-x')^\top}{\sigma^2} \right).
\end{flalign*}
For the score function $s_{p_0}$, we show how to calculate it for some well-known distributions.

\begin{exam}\label{exam_1}
Let $N_d(\mu,\Sigma)$ be the multivariate normal distribution in $\mathbb{R}^{d}$, where $\mu\in\mathbb{R}^{d}$ is the location vector, and $\Sigma\in\mathbb{R}^{d\times d}$ is the scale matrix. When $p_0$ is $N_d(0, \mathrm{I}_d)$, we have $s_{p_0}(x) = -x$.
\end{exam}

\begin{exam}\label{exam_2}
Let $T_d(\mu,\Sigma;\nu)$ be the multivariate $t_{\nu}$ distribution in $\mathbb{R}^{d}$, where $\mu\in\mathbb{R}^{d}$ is the location vector,  $\Sigma\in\mathbb{R}^{d\times d}$ is the scale matrix, and $\nu>2$ is the degrees of freedom. When $p_0$ is $T_d(0, \frac{\nu-2}{\nu}\mathrm{I}_d; \nu)$ (denoted by $T_d(\nu)$) with mean zero and covariance matrix $\mathrm{I}_d$, we have
$$s_{p_0}(x) = - \frac{(\nu + d)x}{\nu -2 + x^\top x}.$$
\end{exam}

\begin{exam}\label{exam_3}
Let $SN_d(\xi, \Omega, \alpha)$ be the multivariate skew-normal distribution in $\mathbb{R}^{d}$ (Arellano-Valle and Azzalini, 2008), where
$\xi\in\mathbb{R}^{d}$ is the location vector, $\Omega\in\mathbb{R}^{d\times d}$ is the scale matrix, and
$\alpha\in\mathbb{R}^{d}$ is the shape vector. To make sure that $SN_d(\xi, \Omega, \alpha)$ has mean zero and covariance matrix $\mathrm{I}_d$, we can
choose the skewness vector $\gamma=(\gamma_1,...,\gamma_{d})^\top\in\mathbb{R}^{d}$ and then set
\begin{equation}\label{E7}
\xi=-  \Sigma_z^{-1}\mu_z, \,\,\, \Omega=\mathrm{I}_d+\xi \xi^\top , \,\,\,\alpha=\dfrac{\big(\bar{\Omega}\big)^{-1}\delta}{\sqrt{1-\delta^\top\big(\bar{\Omega}\big)^{-1}\delta}},
\end{equation}
where
$\Sigma_z=\text{diag}\{\sigma_{z,1},...,\sigma_{z,d}\}$, $\mu_z=(\mu_{z,1},...,\mu_{z,d})^\top$,
$\bar{\Omega}=\Sigma_z \Omega \Sigma_z^{-1}$, and $\delta=\sqrt{\frac{\pi}{2}} \cdot \mu_z$ with
$$\sigma_{z,j}=\big(1-\mu_{z,j}^2\big)^{1/2}, \,\,\, \mu_{z,j}= \dfrac{c_j}{\sqrt{1+c_j^2}}, \,\,\, c_j= \Big(\dfrac{2\gamma_j}{4-\pi}\Big)^{1/3}.$$
Under the settings in (\ref{E7}), we denote $SN_d(\xi, \Omega, \alpha)$ as $SN_d(\gamma)$.
When $p_0$ is $SN_d(\gamma)$  with mean zero and covariance matrix $\mathrm{I}_d$, we have
\begin{equation*} 
s_{p_0}(x) = -\Omega^{-1}(x-\xi)+\dfrac{\phi(\alpha^\top\Sigma_z  (x-\xi))}{\Phi(\alpha^\top\Sigma_z (x-\xi))} \Sigma_z\alpha,
\end{equation*}
where $\phi$ and $\Phi$ denote the $N(0, 1)$ density and distribution functions, respectively.
\end{exam}

Unlike Liu et al. (2016), our test statistic $\widehat{\mathbb{S}}$ does not use the entire data $\{\widehat{\eta}_{t}\}_{t=1}^{n}$.
This is because we have to sacrifice some part of $\{\widehat{\eta}_{t}\}_{t=1}^{n}$ to
deal with the effect of estimation uncertainty
caused by replacing $\theta_0$ via $\widehat{\theta}_{n}$ and the effect of
unobserved initial values resulting from substituting $I_t$ by $\widehat{I}_t$.
With the assist of bootstrap scheme, our numerical studies in Section 4 show that $\widehat{\mathbb{S}}$ can have
good a size and power performance even when no or few data $\{\widehat{\eta}_{t}\}_{t=1}^{n}$ are discarded.
Hence, it suggests that $\widehat{\mathbb{S}}$ can be used with $n_0=n$ or $n_0\approx n$ in practice, and
the subsampling technique seems only theoretically relevant.




\section{Asymptotic theory}

\subsection{Technical assumptions}

Denote
$$g_t(\theta)=g(Y_t, I_{t-1}; \theta),\,\,\,\widehat g_{t}(\theta)=g(Y_{t},\widehat I_{t-1};\theta),\,\,\,\mbox{ and }\,\,\,\widehat{R}_{t}(\theta)=
\widehat g_{t}(\theta)-g_t(\theta),$$
where $g(Y_t, I_{t-1}; \theta)$ is defined in (\ref{E2}).
In this subsection, we give some technical assumptions to study the asymptotics of $\widehat{\mathbb{S}}$.

\begin{assum} \label{asm1}
	$Y_t$ is strictly stationary and ergodic. 
\end{assum}

\begin{assum} \label{asm2}
$E\norm{\eta_t}^4<\infty$.
\end{assum}

\begin{assum} \label{asm3}
	The function $g_t(\theta)$ satisfies that



(i) ${\displaystyle E\Big(\sup_{\theta\in\Theta}\norm{\nabla_{\theta_i} g_t(\theta)}\Big)^2<\infty}$;

(ii) ${\displaystyle E\Big(\sup_{\theta\in\Theta}\norm{\nabla_{\theta_i,\theta_j} g_t(\theta)}\Big)^2<\infty}$,

\noindent for any $i, j \in \{1, ..., p\}$.
\end{assum}

\begin{assum} \label{asm4}
	The estimator $\widehat\theta_n$ satisfies that $\sqrt{n}(\widehat\theta_n-\theta_0)=O_{p}(1)$.
\end{assum}

\begin{assum} \label{asm5}
	The function $\widehat R_{t}(\theta)$ satisfies that
	\begin{equation}
	\sum_{t=1}^\infty E \left(\sup_{\theta}\norm{\widehat R_{t}(\theta)}^4\right)<\infty.\nonumber
	\end{equation}
\end{assum}

\begin{assum} \label{asm6}
The distributions $p$ and $p_0$ satisfy that

(i) both $p$ and $p_0$ are continuous with $\norm{p\delta_{p_0, p}}_2^2<\infty$;


(ii) $\norm{f(x_1)-f(x_2)}<K\norm{x_1-x_2}$, where $f(x)$ is one of $s_{p_0}(x), \nabla_{x_i}s_{p_0}(x)$ and $\nabla_{x_i,x_j}s_{p_0}(x)$,
for any $i, j \in \{1, ..., d\}$, and $K>0$ is a given constant.
\end{assum}

\begin{assum} \label{asm7}
	The kernel function $k(x,x')$ satisfies that

(i) $k(x,x')$ is in the Stein class of $p$;

(ii) $k(x,x')$ and its partial derivatives up to fourth order are all uniformly bounded.
\end{assum}


A few remarks are in order related to the aforementioned assumptions.
Assumptions \ref{asm1}--\ref{asm2} are regular in many time series applications.
Assumption \ref{asm3} poses some moment conditions on the derivatives of $g_t(\theta)$ for the purpose of proof, and
Assumption \ref{asm4} holds for most estimators such as the least squares estimator (LSE) for
VARMA models and the quasi-maximum likelihood estimator (QMLE) for VARMA--GARCH models.
Sufficient conditions to validate  Assumptions \ref{asm3}--\ref{asm4} can be found in
 L\"{u}tkepohl (2005) for VARMA models, Comte and Lieberman (2003), Hafner and Preminger (2009), and Francq and Zako\"{i}an (2012)
 for MGARCH models, and Ling and McAleer (2003) for VARMA--MGARCH models.
Assumption \ref{asm5} is a condition on the approximation error by replacing the
information set $I_t$ by $\widehat{I}_{t}$, and it is used to show that the unobserved initial
values have the negligible effect on the asymptotic theory. See also Hong and Lee (2005) and Escanciano (2006) for the similar conditions.

 Assumption \ref{asm6} requires both $p$ and $p_0$ to have certain smooth conditions.
 The condition $\|p\delta_{p_0,p}\|_{2}^{2}<\infty$ is sufficient to prove the equivalence result (\ref{Equiv}).
 As argued in Liu et al. (2016), this condition is mild. For example, it holds when $p$ is the density function of multivariate normal and $t_{\nu}$ distributions or $p$ has an exponentially decayed tail, but
it may not hold when $p$ has a heavy tail. Note that the exclusion of heavy-tailed $p$ is also implied by Assumption \ref{asm2}.
Assumption \ref{asm7}(i) ensures the validity of (\ref{KSD_2}), and
Assumption \ref{asm7}(ii) poses some boundedness conditions on $k$ and its derivatives.
It is easy to check that
Gaussian kernel in (\ref{Gaussian_k}) satisfies
Assumption \ref{asm7} for any smooth density $p$
supported on $\aleph=\mathbb{R}^d$.
Hence, we follow Liu et al. (2016) to use the Gaussian kernel in this paper.

\subsection{Asymptotics of $\widehat{\mathbb{S}}$}

According to Theorem 3.7 in Liu et al. (2016), the kernel function $u(x,x')$ is positive
definite, and then by Mercer's theorem, $u(x,x')$ admits the expansion
\begin{equation} \label{u_expansion}
u(x,x')=\sum_{m=1}^{\infty}\lambda_m l_{m}(x) l_{m}(x'),
\end{equation}
where $\{l_{m}(\cdot)\}$ and $\{\lambda_m\}$ are the orthonormal eigenfunctions and eigenvalues of $u(x,x')$.
We are ready to give the
limiting null distribution of $\widehat{\mathbb{S}}$.

\begin{thm} \label{thm1}
	Suppose Assumptions \ref{asm1}-\ref{asm7} hold and $\varepsilon>1/2$. Then, under $H_0$,
	$$n_0\widehat{\mathbb{S}}\xrightarrow{d} \chi_0:=\sum_{m=1}^{\infty}\lambda_m(\mathcal{Z}_{m}^2-1) \mbox{ as }n\to\infty,$$
	where $(\mathcal{Z}_{m})_{m\ge 1}$ are i.i.d. standard normal random variables.
\end{thm}

Our limiting null distribution in Theorem \ref{thm1} is the same as the one in Theorem 4.1 of Liu et al. (2016), since
the effects of estimation uncertainty and unobserved initial values are asymptotically negligible by using the sub-sample technique with
$\varepsilon>1/2$. When $\varepsilon\leq1/2$, how to establish the limiting null distribution of $\widehat{\mathbb{S}}$
is unclear at  this stage, and we leave this topic for future study.

Although the effects of estimation uncertainty and unobserved initial values are asymptotically negligible in  theory, they
may exist in finite samples especially when $n$ is small. To redeem this drawback, we propose a simple
parametric bootstrap method in Subsection 3.3 below to calculate the critical values of $n_0\widehat{\mathbb{S}}$. Owing to the use of bootstrap,
our simulation studies will show that $\widehat{\mathbb{S}}$ has a good finite-sample performance even for very small value of
$\varepsilon$, indicating that the condition $\varepsilon>1/2$ should not be an obstacle for applications.

Next, the behavior of $\widehat{\mathbb{S}}$ under $H_1$ is given in the following theorem.

\begin{thm} \label{thm2}
	Suppose Assumptions \ref{asm1}--\ref{asm7} hold. Then, under $H_1$, for any fixed constant $c>0$,
	$$\lim\limits_{n\rightarrow \infty}P(n_0\widehat{\mathbb{S}}>c)=1.$$
\end{thm}

Let $c_{\alpha}$ be the critical value of $n_0\widehat{\mathbb{S}}$ at the level $\alpha$. Then, the preceding theorem implies
that under $H_1$, the power function $\Lambda_{n}:=P(n_0\widehat{\mathbb{S}}>c_{\alpha})$ converges to 1 as $n\to\infty$, and hence
$\widehat{\mathbb{S}}$ can detect $H_1$ consistently.

To end this subsection, we discuss how the choice of $\sigma$ in (\ref{Gaussian_k}) affects the value of $\Lambda_{n}$.
By (\ref{eq_18}) in Appendix A.2, we can show that under $H_1$, for large $n$,
\begin{flalign}\label{3_3}
\Lambda_n&\approx P\Big(\sqrt{n_0}\big(\mathbb{S}^{(0)}-\mathbb{S}(p,p_0)\big)+
\sqrt{n}(\widehat{\theta}_n-\theta_0)^\top\sqrt{\frac{n_0}{n}} \mathbb{S}^{(1)}+\sqrt{n_0}\mathbb{S}(p,p_0)>\frac{c_{\alpha}}{\sqrt{n_0}}\Big).
\end{flalign}
To further calculate $\Lambda_{n}$, we assume $n_0\approx n$ (as recommended for practical use)  and
$$\Big(\sqrt{n}\big(\mathbb{S}^{(0)}-\mathbb{S}(p,p_0)\big), \sqrt{n}(\widehat{\theta}_n-\theta_0)^\top\Big)^\top
\xrightarrow{d} N(0,\Sigma_{\mathbb{S},\theta})$$
as $n\to\infty$, where $\Sigma_{\mathbb{S},\theta}\in\mathbb{R}^{(q+1)\times(q+1)}$ is the asymptotic covariance matrix. Then, by (\ref{3_3}) it is straightforward to see
\begin{flalign}\label{3_4}
\Lambda_n&\approx 1-\Phi\Big(-\frac{\sqrt{n}\mathbb{S}(p,p_0)}{\kappa}\Big) \,\,\,\mbox{ for large }n,
\end{flalign}
where $\kappa=\sqrt{(1,s_1^\top)\Sigma_{\mathbb{S},\theta}(1,s_1^\top)^\top}$,
and $s_1$ is the limit of $\mathbb{S}^{(1)}$ by the law of large numbers for U-statistics.
From (\ref{3_4}), we know that $\sigma$ should be chosen such that $\mathbb{S}(p,p_0)/\kappa$ is maximized. However, this implementation can not
be accomplished in an easy way, since an explicit form of $\kappa$ is not available.
Therefore, it seems hard to choose $\sigma$ optimally. In practice, we can follow Liu et al. (2016) to choose $\sigma$ as the median of residual distance:
 \begin{flalign}\label{sigma_choice}
\sigma = median\{\chi_{ij}, 1 \leq i < j \leq n\},
\end{flalign}
where
$\chi_{ij}={\rVert \widehat{\eta}_i - \widehat{\eta}_j \rVert}^2$.
Our simulation studies in Section 4 below show that $\widehat{\mathbb{S}}$ has a good finite-sample performance based on
this choice of $\sigma$.

\subsection{The computation of critical values}

When $\eta_t$ is observed (i.e., $\eta_t=Y_t$), Liu et al. (2016) applied a Wild bootstrap method to calculate
critical values for their test. However, when $\eta_t$ is unobserved as in our settings,
their bootstrap scheme may not work, since it does not account for the effects of estimation uncertainty and unobserved initial values, which
can affect our critical value $c_{\alpha}$ in the finite sample.
%
%
In this paper, we apply the following parametric bootstrap method to calculate $c_{\alpha}$:

    Step 1. Draw bootstrap i.i.d. errors $\eta_{t}^{*}\sim p_0$ and calculate the bootstrap data sample
    $$Y_{t}^{*}=M(I_{t-1}^{*}; \widehat{\theta}_{n})+C^{1/2}(I_{t-1}^{*}; \widehat{\theta}_{n})\eta_{t}^{*},$$
    where $I_{t-1}^{*}$ is the bootstrap counterpart of  $I_{t-1}$.

    Step 2. Calculate the bootstrap estimator $\widehat{\theta}_{n}^{*}$ and the bootstrap residuals
    $$\widehat{\eta}_{t}^{*}=g(Y_{t}^{*},\widehat{I}_{t-1}^{*};\widehat{\theta}_{n}^{*}),$$
    where $\widehat{I}_{t-1}^{*}$ is the bootstrap counterpart of  $\widehat{I}_{t-1}$.

    Step 3. Compute the bootstrap test statistic $n_0\widehat{\mathbb{S}}^{*}$, based on the bootstrap residuals.

    Step 4. Repeat steps 1--3 $m$ times to get $\{n_0\widehat{\mathbb{S}}_{(1)}^{*},...,n_0\widehat{\mathbb{S}}_{(m)}^{*}\}$,
    whose empirical $\alpha$ upper quantile is taken as the critical value $c_{\alpha}$.

The validity of $c_{\alpha}$ under $H_0$ and $H_1$ can be justified by using the similar arguments as for Theorems \ref{thm1} and \ref{thm2},
respectively, and hence we omit the details.

\section{Simulations}

In this section, we carry out simulation experiments to assess the performance of our KSD-based test $\widehat{\mathbb{S}}$ in finite samples.
For the purpose of comparison, some widely-used tests (see Appendix A.3 for their definitions and asymptotics) are also considered.
The data generating processes (DGPs) considered below cover the dimension $d=2$ and $5$.
In all simulations, we take the sample
size $n=100$ or $500$, choose the number of repetitions $J=10,000$, and
set the significance
level  $\alpha= 1\%$, 5\%, or 10\%.
For $\widehat{\mathbb{S}}$, we use the Gaussian kernel in (\ref{Gaussian_k}) with $\sigma$ taken as in (\ref{sigma_choice}), and choose $n_0=n$ such that no data $\{\widehat{\eta}_t\}_{t=1}^{n}$ are discarded.
To reduce the computational
burden in simulations,
we follow Francq et al. (2017) to adopt the Warp-Speed method of Giacomini et al.
(2013) for evaluating the bootstrap scheme proposed in Subsection 3.3.
With the Warp-Speed method, rather than computing critical
value $c_{\alpha}$ for each repetition sample, only one resample is generated for
each repetition sample and the resampling test statistic $\widehat{\mathbb{S}}^{*}$ is
computed for that sample. Then the critical
value $c_{\alpha}$ is computed from the empirical distribution determined by the
resampling repetitions $\{\widehat{\mathbb{S}}^{*}_{(i)}\}_{i=1}^{J}$.


\subsection{Case 1: Constant mean and constant covariance models}

We consider the DGP given by a constant mean and constant covariance model
\begin{equation}
\label{E4.2.1}
Y_t = M + C^{1/2}\eta_t,
\end{equation}
where $M$ and $C$ are constant mean and constant covariance of $Y_t$, respectively, and they are chosen as
$$M = \left(\begin{array}{c}
    0\\ 0\\
    \end{array} \right), \,\,\,
C^{1/2} = \left( \begin{array}{cc}
    1 & 0.5\\
    0.5 & 1\\
    \end{array} \right)$$
for $d = 2$, and
$$M = \left(\begin{array}{c}
    0\\ 0\\ 0\\ 0\\ 0
    \end{array} \right), \,\,\,
C^{1/2} = \left(\begin{array}{ccccc}
    1 & 0.5 & 0.25 & 0.125 & 0.0625\\
    0.5 & 1 & 0.5 & 0.25 & 0.125 \\
    0.25 & 0.5 & 1 & 0.5 & 0.25 \\
    0.125 & 0.25 & 0.5 & 1 & 0.5 \\
    0.0625 & 0.125 & 0.25 & 0.5 & 1\\
    \end{array} \right)$$
for  $d = 5$.
In model (\ref{E4.2.1}), the distribution of $\eta_t$ (i.e., the true distribution $p$) is $N_d(0,\mathrm{I}_d)$, $T_d(5)$, $SN_d(\gamma)$, or $ST_d(5,\xi)$,
where the first three distributions are given in Examples \ref{exam_1}--\ref{exam_3}, and
the fourth distribution $ST_d(\nu,\xi)$ is the multivariate skew-$t$ distribution in Bauwens and Laurent (2005) with mean zero, covariance matrix $\mathrm{I}_d$,
$\nu>2$ being the degrees of freedom, and
$\xi=(\xi_{1},...,\xi_{d})^\top\in\mathbb{R}^{d}$ being the asymmetry vector to control the skewness.
In the sequel, we set
$$
\gamma=
\left\{
\begin{array}{ll}
(0, -0.6)^\top & \mbox{ for }d=2,\\
(0, 0.2, -0.2, 0, -0.1)^\top & \mbox{ for }d=5,
\end{array}
\right.
\,\,\,
\xi=
\left\{
\begin{array}{ll}
(1, 1.3)^\top & \mbox{ for }d=2,\\
(1.1, 1.2, 1.3, 1.4, 1.5)^\top & \mbox{ for }d=5.
\end{array}
\right.
$$

\begin{table}[!ht]
\caption{Size and power ($\times100$) of all tests in Case 1 for $d=2$}\label{table1}\centering
\small\addtolength{\tabcolsep}{-1.6pt}
 \renewcommand{\arraystretch}{1.1}
\begin{threeparttable}
\begin{tabular}{cccccc|ccc|ccc|ccc}
\hline
      &       &       & \multicolumn{12}{c}{$p(x)$} \bigstrut\\
\cline{4-15}      &       &       & \multicolumn{3}{c|}{$N_d(0, \mathrm{I}_d)$} & \multicolumn{3}{c|}{$T_d(5)$} & \multicolumn{3}{c|}{$SN_{d}(\gamma)$} & \multicolumn{3}{c}{$ST_d(5,\xi)$} \bigstrut\\
\cline{4-15}
$p_0(x)$ & $n$     & Test & 1\%   & 5\%   & 10\%  & 1\%   & 5\%   & 10\%  & 1\%   & 5\%   & 10\%  & 1\%   & 5\%   & 10\% \bigstrut\\
\hline
$N_d(0, \mathrm{I}_d)$      &   100    & $\widehat{\mathbb{S}}$   & 0.6   & 5.5   & 11.5  & 61.0    & 75.9  & 84.1  & 16.8  & 34.5  & 46.3  & 80.7  & 88.9  & 95.2 \bigstrut[t]\\
      &       & $\widehat{\mathbb{T}}_{M,1}$  & 0.7   & 4.6   & 9.5   & 53.3  & 65.5  & 72.0    & 25.7  & 47.0    & 61.1  & 67.4  & 81.9  & 89.3 \\
      &       & $\widehat{\mathbb{T}}_{M,2}$  & 1.4   & 3.2   & 6.9   & 80.6  & 87.3  & 90.6  & 9.1   & 13.1  & 17.0    & 93.4  & 96.0    & 98.8 \\
      &       & $\widehat{\mathbb{T}}_{DH}$    & 1.5   & 5.5   & 10.0    & 70.9  & 82.5  & 87.2  & 24.3  & 42.5  & 53.3  & 84.7  & 91.5  & 96.6 \\
      &       & $\widehat{\mathbb{T}}_{HZ}$    & 0.5   & 5.4   & 11.0    & 48.1  & 66.1  & 75.7  & 9.9   & 26.5  & 38.7  & 73.1  & 84.9  & 92.1 \\
      &       & $\widehat{\mathbb{T}}_{BC,1}$ & 9.3   & 15.5  & 22.7  & 49.7  & 64.2  & 71.5  & 0.2   & 1.4   & 3.3   & 70.3  & 79.4  & 88.2 \\
      &       & $\widehat{\mathbb{T}}_{BC,2}$ & 12.0    & 21.8  & 29.4  & 51.8  & 66.8  & 73.2  & 0.2   & 2.4   & 4.8   & 72.5  & 82.0    & 89.3 \\
      &       & $\widehat{\mathbb{T}}_{BC,3}$ & 11.8  & 19.6  & 27.4  & 53.2  & 66.5  & 74.9  & 0.2   & 1.9   & 5.6   & 73.1  & 84.5  & 90.7 \\
      &       & $\widehat{\mathbb{T}}_{HJM}$ & 0.8   & 5.7   & 11.2  & 56.1  & 68.4  & 77.4  & 4.5   & 12.4  & 20.1  & 75.0    & 86.2  & 92.8 \bigstrut[b]\\
\cline{2-15}  &   500    & $\widehat{\mathbb{S}}$  & 1.3   & 5.3   & 9.6   & 100   & 100   & 100   & 91.9  & 98.2  & 99.1  & 100   & 100   & 100 \bigstrut[t]\\
      &       & $\widehat{\mathbb{T}}_{M,1}$   & 1.2   & 5.6   & 10.1  & 78.4  & 86.1  & 89.6  & 99.0    & 99.9  & 100   & 100   & 100   & 100 \\
      &       & $\widehat{\mathbb{T}}_{M,2}$   & 1.4   & 4.0     & 8.3   & 100   & 100   & 100   & 23.4  & 36.2  & 43.9  & 100   & 100   & 100 \\
      &       & $\widehat{\mathbb{T}}_{DH}$     & 1.2   & 4.0     & 9.5   & 99.9  & 100   & 100   & 98.7  & 99.8  & 99.9  & 100   & 100   & 100 \\
      &       & $\widehat{\mathbb{T}}_{HZ}$     & 0.9   & 4.9   & 9.2   & 99.7  & 100   & 100   & 71.6  & 87.4  & 92.7  & 100   & 100   & 100 \\
      &       & $\widehat{\mathbb{T}}_{BC,1}$ & 8.8   & 14.6  & 23.1  & 88.3  & 93.5  & 97.6  & 0.3   & 2.7   & 10.2  & 100   & 100   & 100 \\
      &       & $\widehat{\mathbb{T}}_{BC,2}$ & 13.2  & 23.5  & 31.7  & 91.2  & 94.1  & 98.2  & 0.6   & 4.5   & 16.4  & 100   & 100   & 100 \\
      &       & $\widehat{\mathbb{T}}_{BC,3}$ & 12.6  & 21.4  & 28.9  & 91.7  & 96.0    & 99.2  & 2.0     & 8.1   & 18.1  & 100   & 100   & 100 \\
      &       & $\widehat{\mathbb{T}}_{HJM}$ & 1.2   & 4.8   & 9.5   & 100   & 100   & 100   & 13.8  & 30.4  & 41.7  & 100   & 100   & 100 \bigstrut[b]\\
\hline
$T_d(5)$      &   100    & $\widehat{\mathbb{S}}$   & 0.4   & 9.4   & 25    & 0.5   & 4.9   & 9.4   & 2     & 26.6  & 48.8  & 4.4   & 20.6  & 33.6 \bigstrut[t]\\
      &       & $\widehat{\mathbb{T}}_{BC,1}$ & 0     & 0     & 0     & 8.5   & 14.2  & 20.1  & 0     & 0     & 4.6   & 2.9   & 9.6   & 18.8 \\
      &       & $\widehat{\mathbb{T}}_{BC,2}$ & 0     & 0     & 0     & 13.5  & 23.7  & 30.6  & 2.3   & 4.1   & 5.3   & 3.6   & 11.2  & 17.7 \\
      &       & $\widehat{\mathbb{T}}_{BC,3}$ & 0     & 0     & 0     & 11.7  & 21.9  & 28.6  & 1.8   & 2.3   & 3.7   & 4.4   & 11.3  & 18.9 \bigstrut[b]\\
\cline{2-15}  &   500    & $\widehat{\mathbb{S}}$   & 81.8  & 99.5  & 100   & 0.4   & 4.6   & 10.5  & 99.4  & 100   & 100   & 90.6  & 100   & 100 \bigstrut[t]\\
      &       & $\widehat{\mathbb{T}}_{BC,1}$ & 0     & 0     & 0.4   & 7.8   & 14.5  & 22.7  & 0.5   & 4.3   & 12.1  & 77.3  & 96.3  & 100 \\
      &       & $\widehat{\mathbb{T}}_{BC,2}$ & 0     & 0.1   & 1.1   & 13.1  & 23.6  & 31.5  & 1.4   & 6.7   & 16.8  & 79.1  & 96.9  & 100 \\
      &       & $\widehat{\mathbb{T}}_{BC,3}$ & 0     & 0     & 0.8   & 12.2  & 20.4  & 30.2  & 1.4   & 7.1   & 18.5  & 80.7  & 98.5  & 100 \bigstrut[b]\\
\hline
$SN_d(\gamma)$ & 100   & $\widehat{\mathbb{S}}$  & 14.9  & 32.5  & 45.1  & 73.6  & 87.7  & 91.9  & 0.9   & 5.0     & 10.2  & 94.6  & 98.0    & 98.7 \bigstrut[t]\\
      & 500   & $\widehat{\mathbb{S}}$   & 95.9  & 99.3  & 99.9  & 100   & 100   & 100   & 0.5   & 3.9   & 8.4   & 100   & 100   & 100 \bigstrut[b]\\
\hline
\end{tabular}%
\end{threeparttable}
\end{table}

\begin{table}[!ht]
\caption{Size and power ($\times100$) of all tests in Case 1 for $d=5$}\label{table2}\centering
\small\addtolength{\tabcolsep}{-1.6pt}
 \renewcommand{\arraystretch}{1.1}
\begin{threeparttable}

\begin{tabular}{cccccc|ccc|ccc|ccc}
\hline
      &       &       & \multicolumn{12}{c}{$p(x)$} \bigstrut\\
\cline{4-15}      &       &       & \multicolumn{3}{c|}{$N_d(0, \mathrm{I}_d$)} & \multicolumn{3}{c|}{$T_{d}(5)$} & \multicolumn{3}{c|}{$SN_d(\gamma)$} & \multicolumn{3}{c}{$ST_{d}(5,\xi)$} \bigstrut\\
\cline{4-15}
$p_0(x)$ & $n$     & Test & 1\%   & 5\%   & 10\%  & 1\%   & 5\%   & 10\%  & 1\%   & 5\%   & 10\%  & 1\%   & 5\%   & 10\% \bigstrut\\
\hline
$N_d(0, \mathrm{I}_d)$      &   100    & $\widehat{\mathbb{S}}$   & 1.4   & 4.8   & 9.7   & 97.9  & 99.5  & 99.9  & 17.0    & 38.5  & 52.3  & 100   & 100   & 100 \bigstrut[t]\\
      &       & $\widehat{\mathbb{T}}_{M,1}$  & 1.1   & 4.7   & 8.4   & 95.2  & 97.3  & 98.4  & 3.2   & 10.9  & 16.7  & 100   & 100   & 100 \\
      &       & $\widehat{\mathbb{T}}_{M,2}$  & 0.5   & 2.7   & 8.5   & 99.5  & 100   & 100   & 3.8   & 7.3   & 10.5  & 100   & 100   & 100 \\
      &       & $\widehat{\mathbb{T}}_{DH}$    & 1.6   & 5.1   & 9.7   & 90.8  & 95.6  & 97.3  & 3.3   & 10.8  & 18.2  & 100   & 100   & 100 \\
      &       & $\widehat{\mathbb{T}}_{HZ}$    & 0.8   & 4.8   & 11    & 92.9  & 96.4  & 98    & 7.8   & 20.7  & 30.5  & 100   & 100   & 100 \\
      &       & $\widehat{\mathbb{T}}_{HJM}$ & 1.4   & 5.2   & 10.4  & 94.3  & 98.7  & 99.3  & 3.2   & 10.1  & 17.4  & 100   & 100   & 100 \bigstrut[b]\\
\cline{2-15}  &   500    & $\widehat{\mathbb{S}}$   & 1.5   & 5.0   & 9.3   & 100   & 100   & 100   & 99.1  & 99.9  & 100   & 100   & 100   & 100  \bigstrut[t]\\
      &       & $\widehat{\mathbb{T}}_{M,1}$  & 1.1   & 4.8   & 10.2  & 100   & 100   & 100   & 27.9  & 50.0    & 64.0    & 100   & 100   & 100 \\
      &       & $\widehat{\mathbb{T}}_{M,2}$  & 0.4   & 4.4   & 8.4   & 100   & 100   & 100   & 5.3   & 10.6  & 16.6  & 100   & 100   & 100 \\
      &       & $\widehat{\mathbb{T}}_{DH}$    & 1.3   & 5.1   & 9.7   & 100   & 100   & 100   & 19.5  & 38.1  & 50.8  & 100   & 100   & 100 \\
      &       & $\widehat{\mathbb{T}}_{HZ}$    & 1.0   & 5.0   & 8.2   & 100   & 100   & 100   & 72.1  & 89.3  & 93.6  & 100   & 100   & 100 \\
      &       & $\widehat{\mathbb{T}}_{HJM}$ & 1.4   & 5.2   & 8.9   & 100   & 100   & 100   & 9.0   & 22.9  & 33.8  & 100   & 100   & 100 \bigstrut[b]\\
\hline
$T_d(5)$ & 100   & $\widehat{\mathbb{S}}$   & 4.2   & 15.8  & 38.1  & 1.3   & 4.9   & 9.8   & 0     & 0.1   & 0.4   & 26.8  & 50.3  & 64.9 \bigstrut\\
\cline{2-15}      & 500   & $\widehat{\mathbb{S}}$   & 87.3  & 100   & 100   & 1.0     & 4.7   & 10.2  & 75.3  & 100   & 100   & 100   & 100   & 100 \bigstrut\\
\hline
$SN_d(\gamma)$ & 100   & $\widehat{\mathbb{S}}$   & 60.9  & 75.1  & 80.5  & 89.5  & 96.7  & 98.5  & 1.0     & 4.2   & 8.7   & 99.2  & 99.8  & 99.9 \bigstrut\\
\cline{2-15}      & 500   & $\widehat{\mathbb{S}}$   & 100   & 100   & 100   & 100   & 100   & 100   & 1.1   & 4.9   & 9.7   & 100   & 100   & 100 \bigstrut\\
\hline
\end{tabular}%

\end{threeparttable}
\end{table}

For the null distribution $p_0$ in (\ref{Hypothesis}), we take it to be $N_d(0, \mathrm{I}_d)$, $T_d(5)$, or $SN_d(\gamma)$.
When $p_0$ is $N_d(0, \mathrm{I}_d)$, we also consider Mardia's skewness test ($\widehat{\mathbb{T}}_{M,1}$), Mardia's kurtosis test ($\widehat{\mathbb{T}}_{M,2}$),
Doornik--Hansen test ($\widehat{\mathbb{T}}_{DH}$), Henze--Zirkler test ($\widehat{\mathbb{T}}_{HZ}$), Bai--Chen tests ($\widehat{\mathbb{T}}_{BC,1}$, $\widehat{\mathbb{T}}_{BC,2}$, and $\widehat{\mathbb{T}}_{BC,3}$), and
Henze--Jim\'{e}nez-Gamero--Meintanis test ($\widehat{\mathbb{T}}_{HJM}$).
The first four tests $\widehat{\mathbb{T}}_{M,1}$, $\widehat{\mathbb{T}}_{M,2}$, $\widehat{\mathbb{T}}_{DH}$, and $\widehat{\mathbb{T}}_{HZ}$ are based on either the sample skewness or the sample kurtosis or both.
The tests $\widehat{\mathbb{T}}_{BC,i}$, $i=1,2,3$, are based on the empirical distribution of the residuals, and the test
$\widehat{\mathbb{T}}_{HJM}$ is based on the characteristic function of the residuals.
Note that except $\widehat{\mathbb{T}}_{BC,i}$, all other tests work for $d>2$.
When $p_0$ is $T_d(5)$ or $SN_d(\gamma)$, none of the competitive tests above is applicable, except that
the tests $\widehat{\mathbb{T}}_{BC,i}$ can be used for the case of $T_2(5)$.

Tables \ref{table1} and \ref{table2} report the size and power of all examined tests for $d=2$ and $d=5$, respectively, where
the size corresponds to the case of $p=p_0$. In calculation of $\widehat{\mathbb{S}}$, $\widehat{\mathbb{T}}_{BC,i}$, and
$\widehat{\mathbb{T}}_{HJM}$, the residuals of model (\ref{E4.2.1}) are computed by estimating $M$ and $C$ by the sample mean and sample covariance of $Y_t$, respectively.
Note that since the tests $\widehat{\mathbb{T}}_{BC,i}$ are largely over-sized,
we compute their size-adjusted power in the sequel. From Tables \ref{table1} and \ref{table2}, our findings are as follows:

(1) Except for the tests $\widehat{\mathbb{T}}_{BC,i}$, all examined tests have an accurate size performance at three levels.

(2) When $p_0$ is $N_d(0, \mathrm{I}_d)$, $\widehat{\mathbb{S}}$ has a comparative power performance with any competitive test to detect the
 alternative hypotheses that $p$ are $SN_d(\gamma)$ and $ST_d(5,\xi)$. However, $\widehat{\mathbb{S}}$ has the best power performance
 to detect the alternative hypothesis that $p$ is $SN_{d}(\gamma)$, and
 the tests $\widehat{\mathbb{T}}_{M,2}$, $\widehat{\mathbb{T}}_{BC,i}$, and $\widehat{\mathbb{T}}_{HJM}$ have a much worse power performance in this case.
 The advantage of $\widehat{\mathbb{S}}$ is more obvious for the case $d=5$.

 (3) When $p_0$ is $T_d(5)$, $\widehat{\mathbb{S}}$ has the satisfactory power performance especially for $n=500$, while the tests $\widehat{\mathbb{T}}_{BC,i}$ only
 exhibit the power to detect the alternative hypothesis that $p$ is $ST_d(5,\xi)$.

 (4) When $p_0$ is $SN_d(\gamma)$, $\widehat{\mathbb{S}}$ is powerful to detect each examined alternative hypothesis, and its power to detect the
 heavy-tailed alternative distribution (e.g., $T_d(5)$ or $ST_d(5,\xi)$) is higher than that to detect the light-tailed alternative distribution (e.g., $N_d(0, \mathrm{I}_d)$).

 Overall, our KSD-based test $\widehat{\mathbb{S}}$ exhibits the good size and power performance in all examined cases.
All skewness- or kurtosis-based tests for normality generally perform well, except that $\widehat{\mathbb{T}}_{M,2}$ lacks the power to
detect the alternative distribution $SN_{d}(\gamma)$. The tests $\widehat{\mathbb{T}}_{BC,i}$ have the over-sized problem in all examined cases, and their size-adjusted power
in general is not satisfactory especially for the null distribution $T_{2}(5)$.
The test $\widehat{\mathbb{T}}_{HJM}$ for the normality performs as good as $\widehat{\mathbb{S}}$, except that its power to detect
the alternative distribution $SN_d(\gamma)$ is lower. Based on the aforementioned findings, it is reasonable to recommend
$\widehat{\mathbb{S}}$ for use due to its generality and desirable power performance.

\subsection{Case 2: VAR models}

We consider the DGP given by a VAR(3) model
\begin{equation}\label{E4.2.2}
    Y_t = M+A_1Y_{t-1} + A_2Y_{t-2} + A_3Y_{t-3} + C^{1/2}\eta_t,
\end{equation}
where $M$, $C^{1/2}$, and $\eta_t$ are chosen as in model (\ref{E4.2.1}), and
$$A_1 = \begin{pmatrix}
    0.3 & 0.65\\
    -0.2 & -0.4\\
    \end{pmatrix},\quad
  A_2 =\begin{pmatrix}
    -0.4 & 0.4\\
    -0.6 & 0.4\\
    \end{pmatrix},\quad
  A_3 = \begin{pmatrix}
    0.5 & 0.1\\
    0.1 & 0.5\\
    \end{pmatrix}
$$
for $d = 2$, and
\begin{flalign*}
A_1 &= \begin{pmatrix}
    0.2 & 0.1 & -0.2 & 0 & 0\\
    0 & -0.3 & 0.1 & -0.1 & 0\\
    0& 0.05 & 0.15 & 0 & 0 \\
    -0.05 & 0 & 0.1 & -0.2 & 0\\
    0.05 & -0.1 & -0.1 & 0 & 0.3\\
    \end{pmatrix},\,\,
  A_2 = \begin{pmatrix}
    0.25 & 0.05 & 0.1 & 0 & 0\\
    -0.2 & 0.1 & 0.1 & 0 & 0 \\
    0.1 & 0.1 & -0.2 & 0 & 0\\
    0 & 0 & 0 & -0.1 & 0.1 \\
    0 & 0 & 0 & 0.2 & 0.3\\
    \end{pmatrix},\\
  A_3 &= \begin{pmatrix}
    -0.3 & 0.05 &0.1 & 0 & 0 \\
    -0.2 & 0.2 & 0.1 & 0 & 0\\
    0.05 & -0.1 & 0.2 & 0 & 0 \\
    0 & 0 & 0 & -0.15 & -0.1\\
    0 & 0 & 0 & 0.05 & 0.2\\
    \end{pmatrix}
\end{flalign*}
for $d = 5$. As in Case 1, the null distribution $p_0$ is $N_d(0,\mathrm{I}_d)$, $T_d(5)$, or $SN_d(\gamma)$.
For the VAR(3) model in (\ref{E4.2.2}), the skewness- or kurtosis-based tests considered in Case 1 are not applicable any more.
In this case, the tests $\widehat{\mathbb{T}}_{BC,i}$ work when $p_0$ is $N_d(0,\mathrm{I}_d)$ or $T_d(5)$ for $d=2$, and
the test $\widehat{\mathbb{T}}_{HJM}$ works when $p_0$ is $N_d(0,\mathrm{I}_d)$ for $d=2$ and 5.

Tables \ref{table3} and \ref{table4} report the size and power of all examined tests for $d=2$ and $d=5$, respectively, where
the size corresponds to the case of $p=p_0$. In calculation of $\widehat{\mathbb{S}}$, $\widehat{\mathbb{T}}_{BC,i}$, and
$\widehat{\mathbb{T}}_{HJM}$, the residuals of model (\ref{E4.2.2}) are computed by using the LSE to estimate the unknown parameters.
From Tables \ref{table3} and \ref{table4}, our findings are similar as those in Case 1.

\begin{table}[!ht]
\caption{Size and power ($\times100$) of all tests in Case 2 for $d=2$}\label{table3}\centering
\small\addtolength{\tabcolsep}{-1.5pt}
 \renewcommand{\arraystretch}{1.1}
\begin{threeparttable}

\begin{tabular}{cccccc|ccc|ccc|ccc}
\hline
      &       &       & \multicolumn{12}{c}{$p(x)$} \bigstrut\\
\cline{4-15}      &       &       & \multicolumn{3}{c|}{$N_d(0,\mathrm{I}_d)$} & \multicolumn{3}{c|}{$T_d(5)$} & \multicolumn{3}{c|}{$SN_d(\gamma)$} & \multicolumn{3}{c}{$ST_d(5,\xi)$} \bigstrut\\
\cline{4-15}$p_0(x)$ & $n$     & Test & 1\%   & 5\%   & 10\%  & 1\%   & 5\%   & 10\%  & 1\%   & 5\%   & 10\%  & 1\%   & 5\%   & 10\% \bigstrut\\
\hline
$N_d(0,\mathrm{I}_d)$   &   100    & $\widehat{\mathbb{S}}$   & 1.7   & 5.7   & 10.1  & 52.3  & 70.4  & 78.6  & 11.4  & 28.9  & 40.2  & 61.5  & 77.3  & 87.2 \bigstrut[t]\\
      &       & $\widehat{\mathbb{T}}_{BC,1}$ & 7.3   & 12.6  & 18.7  & 43.6  & 56.8  & 64.2  & 0     & 0.7   & 2.1   & 47.2  & 63.3  & 71.6 \\
      &       & $\widehat{\mathbb{T}}_{BC,2}$ & 11.8  & 19.5  & 26.1  & 46.2  & 57.4  & 68.2  & 0     & 1.5   & 3.6   & 50.2  & 62.9  & 77.5 \\
      &       & $\widehat{\mathbb{T}}_{BC,3}$ & 11.4  & 20.7  & 25.0    & 46.9  & 59.0    & 69.3  & 0     & 1.2   & 3.8   & 51.0    & 65.2  & 79.3 \\
      &       & $\widehat{\mathbb{T}}_{HJM}$ & 0.5   & 4.1   & 11.3  & 55.0    & 64.7  & 72.5  & 3.9   & 12.5  & 20.5  & 58.3  & 71.8  & 84.9 \bigstrut[b]\\
\cline{2-15}  &   500    & $\widehat{\mathbb{S}}$   & 0.6   & 4.4   & 9.5   & 99.9  & 100   & 100   & 91.2  & 97.4  & 98.6  & 100   & 100   & 100 \bigstrut[t]\\
      &       & $\widehat{\mathbb{T}}_{BC,1}$ & 8.5   & 13.2  & 19.0    & 85.2  & 90.4  & 98.4  & 0.2   & 2.4   & 10.5  & 100   & 100   & 100 \\
      &       & $\widehat{\mathbb{T}}_{BC,2}$ & 13.5  & 22.2  & 28.5  & 86.0    & 93.1  & 99.0    & 0.6   & 4.1   & 15.7  & 100   & 100   & 100 \\
      &       & $\widehat{\mathbb{T}}_{BC,3}$ & 11.7  & 19.4  & 26.1  & 88.2  & 93.8  & 99.3  & 1.8   & 8.0     & 17.5  & 100   & 100   & 100 \\
      &       & $\widehat{\mathbb{T}}_{HJM}$ & 0.8   & 5.4   & 10.4  & 100   & 100   & 100   & 13.0    & 29.9  & 41.5  & 100   & 100   & 100 \bigstrut[b]\\
\hline
$T_d(5)$       &   100    & $\widehat{\mathbb{S}}$   & 0.2   & 11.1  & 26.9  & 0.8   & 5.7   & 11.0    & 1.6   & 23.7  & 45.9  & 3.7   & 16.6  & 26.1 \bigstrut[t]\\
      &       & $\widehat{\mathbb{T}}_{BC,1}$ & 0     & 0.2   & 0.2   & 9.4   & 15.7  & 23.2  & 0.5   & 1.4   & 3.5   & 3.4   & 7.8   & 14.8 \\
      &       & $\widehat{\mathbb{T}}_{BC,2}$ & 0     & 0.3   & 0.4   & 12.4  & 24.0    & 33.5  & 1.7   & 3.8   & 4.9   & 4.2   & 8.3   & 13.1 \\
      &       & $\widehat{\mathbb{T}}_{BC,3}$ & 0     & 0.3   & 0.4   & 10.9  & 21.6  & 30.6  & 1.1   & 2.4   & 3.7   & 4.6   & 8.2   & 15.7 \bigstrut[b]\\
\cline{2-15}  &   500    & $\widehat{\mathbb{S}}$   & 88.0    & 99.3  & 99.9  & 0.7   & 4.8   & 9.9   & 98.1  & 99.4  & 100   & 84.4  & 95.1  & 99.7  \bigstrut[t]\\
      &       & $\widehat{\mathbb{T}}_{BC,1}$ & 0     & 0     & 0.6   & 8.9   & 14.1  & 20.4  & 0.7   & 3.9   & 11.4  & 63.9  & 75.5  & 84.5 \\
      &       & $\widehat{\mathbb{T}}_{BC,2}$ & 0     & 0.4   & 1.4   & 12.7  & 21.5  & 29.3  & 1.5   & 5.8   & 16.2  & 65.3  & 74.1  & 85.9 \\
      &       & $\widehat{\mathbb{T}}_{BC,3}$ & 0     & 0.4   & 1.2   & 11.4  & 18.6  & 26.7  & 1.4   & 6.5   & 17.5  & 66.0    & 78.2  & 89.3 \bigstrut[b]\\
\hline
$SN_d(\gamma)$ & 100   & $\widehat{\mathbb{S}}$   & 14.8  & 34.4  & 45.3  & 67.4  & 81.5  & 88.3  & 1.0     & 6.1   & 9.9   & 90.7  & 95.2  & 98.1 \bigstrut\\
\cline{2-15}      & 500   & $\widehat{\mathbb{S}}$   & 95.4  & 99.2  & 100   & 100   & 100   & 100   & 0.8   & 5.3   & 9.8   & 100   & 100   & 100 \bigstrut\\
\hline
\end{tabular}%

\end{threeparttable}
\end{table}

\begin{table}[!ht]
\caption{Size and power ($\times100$) of all tests in Case 2 for $d=5$}\label{table4}\centering
\small\addtolength{\tabcolsep}{-1.5pt}
 \renewcommand{\arraystretch}{1.1}
\begin{threeparttable}

\begin{tabular}{cccccc|ccc|ccc|ccc}
\hline
      &       &       & \multicolumn{12}{c}{$p(x)$} \bigstrut\\
\cline{4-15}      &       &       & \multicolumn{3}{c|}{$N_d(0, \mathrm{I}_d)$} & \multicolumn{3}{c|}{$T_d(5)$} & \multicolumn{3}{c|}{$SN_d(\gamma)$} & \multicolumn{3}{c}{$ST_d(\xi)$} \bigstrut\\
\cline{4-15}$p_0(x)$ & $n$     & Test & 1\%   & 5\%   & 10\%  & 1\%   & 5\%   & 10\%  & 1\%   & 5\%   & 10\%  & 1\%   & 5\%   & 10\% \bigstrut\\
\hline
$N_d(0, \mathrm{I}_d)$ &   100    & $\widehat{\mathbb{S}}$   & 1.1   & 4.8   & 9.6   & 89.4  & 96.7  & 99.1  & 12.1  & 29.7  & 42.1  & 98.4  & 100   & 100 \bigstrut[t]\\
      &       & $\widehat{\mathbb{T}}_{HJM}$ & 0.6   & 4.7   & 10.3  & .0    & 95.2  & 98.3  & 3.7   & 11.8  & 18.9  & 97.7  & 99.7  & 100 \bigstrut[b]\\
\cline{2-15}  &   500    & $\widehat{\mathbb{S}}$   & 1.3   & 5.2   & 10.7  & 100   & 100   & 100   & 98.5  & 99.6  & 100   & 100   & 100   & 100 \bigstrut[t]\\
      &       & $\widehat{\mathbb{T}}_{HJM}$ & 0.9   & 4.7   & 9.2   & 100   & 100   & 100   & 9.3   & 23.4  & 32.8  & 100   & 100   & 100 \bigstrut[b]\\
\hline
$T_d(5)$ & 100   & $\widehat{\mathbb{S}}$   & 2.7   & 11.8  & 36.4  & 1.4   & 6.2   & 10.5  & 0     & 0.2   & 0.6   & 25.3  & 48.2  & 59.7 \bigstrut\\
\cline{2-15}      & 500   & $\widehat{\mathbb{S}}$   & 100   & 100   & 100   & 1.4   & 5.7   & 11.1  & 72.8  & 99.2  & 100   & 100   & 100   & 100 \bigstrut\\
\hline
$SN_d(\gamma)$ & 100   & $\widehat{\mathbb{S}}$   & 58.4  & 73.8  & 78.8  & 84.7  & 91.4  & 95.3  & 0.7   & 4.5   & 10.4  & 95.8  & 98.5  & 99.7 \bigstrut\\
\cline{2-15}      & 500   & $\widehat{\mathbb{S}}$   & 100   & 100   & 100   & 100   & 100   & 100   & 0.9   & 5.8   & 10.3  & 100   & 100   & 100 \bigstrut\\
\hline
\end{tabular}%

\end{threeparttable}
\end{table}

\subsection{Case 3: CCC--GARCH models}

We consider the DGP given by a CCC-GARCH(1, 1) model
\begin{flalign}\label{E4.2.3}
Y_t
= C_t^{\frac{1}{2}}\eta_t,
\end{flalign}
where $\eta_t$ is chosen as in model (\ref{E4.2.1}), and $C_t=\mbox{diag}\{\sigma_{1,t},...,\sigma_{d,t}\} \cdot R\cdot \mbox{diag}\{\sigma_{1,t},...,\sigma_{d,t}\}$ with
\begin{flalign*}
\begin{pmatrix}
    \sigma_{1,t}^2 \\
    \sigma_{2,t}^2 \\
    \vdots \\
    \sigma_{d,t}^2
  \end{pmatrix}
  =  W +
    B
    \begin{pmatrix}
        Y_{1,t-1}^2\\
        Y_{2,t-1}^2\\
        \vdots\\
        Y_{d,t-1}^2
    \end{pmatrix} +
    \Gamma
\begin{pmatrix}
    \sigma_{1,t-1}^2 \\
    \sigma_{2,t-1}^2 \\
    \vdots \\
    \sigma_{d,t-1}^2
  \end{pmatrix}.
\end{flalign*}
Here, the parameter matrices $R$, $W$, $B$, and $\Gamma$ are set to be
\begin{flalign*}
R = \begin{pmatrix}
    1 & 0.5\\
    0.5 & 1
\end{pmatrix},\,\,
W  =   \begin{pmatrix}
        0.1\\
        0.1
    \end{pmatrix}, \,\,
B  =
    \begin{pmatrix}
        0.3 & 0.1\\
        0.1 & 0.2
    \end{pmatrix},\,\,
\Gamma
    = \begin{pmatrix}
        0.2 & 0.01\\
        0.1 & 0.3
    \end{pmatrix}
\end{flalign*}
for $d=2$, and
\begin{flalign*}
R &= \begin{pmatrix}
    1 & 0.5 & 0.5 & 0.5 & 0.5\\
    0.5 & 1 & 0.5 & 0.5 & 0.5\\
    0.5 & 0.5 & 1 & 0.5 & 0.5\\
    0.5 & 0.5 & 0.5 & 1 & 0.5\\
    0.5 & 0.5 & 0.5 & 0.5 & 1
\end{pmatrix},\,\,
W
=   \begin{pmatrix}
        0.1\\
        0.1\\
        0.1\\
        0.1\\
        0.1
    \end{pmatrix}, \\
B &=
    \begin{pmatrix}
        0.3 & 0.1 & 0.1 & 0.1 & 0.1\\
        0.1 & 0.2 & 0.1 & 0.1 & 0.1\\
        0.1 & 0.1 & 0.25 & 0.1 & 0.1\\
        0.1 & 0.1 & 0.1 & 0.15 & 0.1\\
        0.1 & 0.1 & 0.1 & 0.1 & 0.1
    \end{pmatrix},\,\,
\Gamma = \begin{pmatrix}
        0.2 & 0.01 & 0.01 & 0.1 & 0.01\\
        0.1 & 0.3 & 0.01 & 0.01 & 0.01\\
        0.01 & 0.1 & 0.1 & 0.01 & 0.1\\
        0.1 & 0.1 & 0.1 & 0.15 & 0.01\\
        0.01 & 0.01 & 0.01 & 0.1 & 0.2
    \end{pmatrix}
\end{flalign*}
for $d=5$.

As in Cases 1 and 2, the null distribution $p_0$ is $N_d(0,\mathrm{I}_d)$, $T_d(5)$, or $SN_d(\gamma)$.
For the CCC-GARCH model in (\ref{E4.2.3}), the competitive tests can only be chosen as in Case 2.
Tables \ref{table5} and \ref{table6} report the size and power of all examined tests for $d=2$ and $d=5$, respectively, where
the size corresponds to the case of $p=p_0$. In calculation of $\widehat{\mathbb{S}}$, $\widehat{\mathbb{T}}_{BC,i}$, and
$\widehat{\mathbb{T}}_{HJM}$, the residuals of model (\ref{E4.2.3}) are computed by using the QMLE to estimate the unknown parameters.
Clearly, our findings from Tables \ref{table5} and \ref{table6} are similar to those in Case 1.

\begin{table}[!ht]
\caption{Size and power ($\times100$) of all tests in Case 3 for $d=2$}\label{table5}\centering
\small\addtolength{\tabcolsep}{-1.5pt}
 \renewcommand{\arraystretch}{1.1}
\begin{threeparttable}

\begin{tabular}{cccccc|ccc|ccc|ccc}
\hline
      &       &       & \multicolumn{12}{c}{$p(x)$} \bigstrut\\
\cline{4-15}      &       &       & \multicolumn{3}{c|}{$N_d(0, \mathrm{I}_d)$} & \multicolumn{3}{c|}{$T_d(5)$} & \multicolumn{3}{c|}{$SN_d(\gamma)$} & \multicolumn{3}{c}{$ST_d(5,\xi)$} \bigstrut\\
\cline{4-15}\multicolumn{1}{c}{$p_0(x)$} & \multicolumn{1}{c}{$n$} & \multicolumn{1}{c}{Test} & \multicolumn{1}{c}{1\%} & \multicolumn{1}{c}{5\%} & \multicolumn{1}{c|}{10\%} & \multicolumn{1}{c}{1\%} & \multicolumn{1}{c}{5\%} & \multicolumn{1}{c|}{10\%} & \multicolumn{1}{c}{1\%} & \multicolumn{1}{c}{5\%} & \multicolumn{1}{c|}{10\%} & \multicolumn{1}{c}{1\%} & \multicolumn{1}{c}{5\%} & \multicolumn{1}{c}{10\%} \bigstrut\\
\hline
$N_d(0, \mathrm{I}_d)$       &   500    & $\widehat{\mathbb{S}}$   & 0.9   & 5.0     & 10.3  & 97.3  & 99.7  & 100   & 89.5  & 96.2  & 98.0    & 100   & 100   & 100 \bigstrut[t]\\
      &       & $\widehat{\mathbb{T}}_{BC,1}$ & 10.4  & 16.1  & 22.5  & 85.1  & 89.9  & 95.2  & 0     & 1.5   & 6.5   & 100   & 100   & 100 \\
      &       & $\widehat{\mathbb{T}}_{BC,2}$ & 14.3  & 23.0    & 31.6  & 88.3  & 93.0    & 96.7  & 0.7   & 3.2   & 13.1  & 100   & 100   & 100 \\
      &       & $\widehat{\mathbb{T}}_{BC,3}$ & 12.7  & 20.4  & 28.3  & 90.4  & 94.1  & 98.4  & 1.6   & 5.7   & 14.9  & 100   & 100   & 100 \\
      &       & $\widehat{\mathbb{T}}_{HJM}$ & 0.8   & 4.7   & 9.7   & 87.7  & 91.4  & 96.2  & 9.2   & 23.7  & 33.5  & 100   & 100   & 100 \bigstrut[b]\\
\hline
$T_d(5)$       &   500    & $\widehat{\mathbb{S}}$   & 92.4  & 99.5  & 99.8  & 0.7   & 5.5   & 10.9  & 97.5  & 99.2  & 100   & 100   & 100   & 100 \bigstrut[t]\\
      &       & $\widehat{\mathbb{T}}_{BC,1}$ & 0     & 0     & 0.6   & 8.9   & 13.2  & 18.5  & 0.6   & 3.4   & 10.1  & 100   & 100   & 100 \\
      &       & $\widehat{\mathbb{T}}_{BC,2}$ & 0.2   & 0.2   & 1.4   & 11.4  & 17.9  & 25.7  & 1.3   & 5.2   & 14.4  & 100   & 100   & 100 \\
      &       & $\widehat{\mathbb{T}}_{BC,3}$ & 0.2   & 0.3   & 1.7   & 9.6   & 17.1  & 24.9  & 1.3   & 5.8   & 15.7  & 100   & 100   & 100 \bigstrut[b]\\
\hline
$SN_{d}(\gamma)$ & 500   & $\widehat{\mathbb{S}}$   & 93.5  & 97.4  & 99.4  & 100   & 100   & 100   & 1.2   & 5.5   & 10.6  & 100   & 100   & 100 \bigstrut\\
\hline
\end{tabular}%


\end{threeparttable}

\end{table}

\begin{table}[!htp]
\caption{Size and power ($\times100$) of all tests in Case 3 for $d=5$}\label{table6}\centering
\small\addtolength{\tabcolsep}{-1.2pt}
 \renewcommand{\arraystretch}{1.1}
\begin{threeparttable}

\begin{tabular}{cccccc|ccc|ccc|ccc}
\hline
      &       &       & \multicolumn{12}{c}{$p(x)$} \bigstrut\\
\cline{4-15}      &       &       & \multicolumn{3}{c|}{$N_d(0, \mathrm{I}_d)$} & \multicolumn{3}{c|}{$T_d(5)$} & \multicolumn{3}{c|}{$SN_d(\gamma)$} & \multicolumn{3}{c}{$ST_d(5,\xi)$} \bigstrut\\
\cline{4-15}\multicolumn{1}{c}{$p_0(x)$} & \multicolumn{1}{c}{$n$} & \multicolumn{1}{c}{Test} & \multicolumn{1}{c}{1\%} & \multicolumn{1}{c}{5\%} & \multicolumn{1}{c|}{10\%} & \multicolumn{1}{c}{1\%} & \multicolumn{1}{c}{5\%} & \multicolumn{1}{c|}{10\%} & \multicolumn{1}{c}{1\%} & \multicolumn{1}{c}{5\%} & \multicolumn{1}{c|}{10\%} & \multicolumn{1}{c}{1\%} & \multicolumn{1}{c}{5\%} & \multicolumn{1}{c}{10\%} \bigstrut\\
\hline
$N_d(0, \mathrm{I}_d)$       &   500    & $\widehat{\mathbb{S}}$   & 1.3   & 5.6   & 11.2  & 100   & 100   & 100   & 96.8  & 98.3  & 99.7  & 100   & 100   & 100  \bigstrut[t]\\
&    & $\widehat{\mathbb{T}}_{HJM}$ & 0.7   & 4.6   & 9.5   & 100   & 100   & 100   & 8.5   & 22.8  & 30.6  & 100   & 100   & 100\bigstrut[b]\\
\hline
$T_d(5)$ & 500   & $\widehat{\mathbb{S}}$   & 100   & 100   & 100   & 1.2   & 5.1   & 10.4  & 68.7  & 98.4  & 100   & 100   & 100   & 100 \bigstrut\\
\hline
$SN_d(\gamma)$ & 500   & $\widehat{\mathbb{S}}$   & 100   & 100   & 100   & 100   & 100   & 100   & 1.3   & 4.8   & 10.5  & 100   & 100   & 100 \bigstrut\\
\hline
\end{tabular}%


\end{threeparttable}
\end{table}

\subsection{Sensitivity analysis}
In our previous simulation studies, we take $n_0=n$ and $\sigma$ as in (\ref{sigma_choice}) to compute our KSD-based test $\widehat{\mathbb{S}}$.
In this subsection, we implement the sensitivity analysis on the choice of $n_0$ or $\sigma$ for $\widehat{\mathbb{S}}$, based on the DGP in (\ref{E4.2.1})
with $p$ being $N_d(0, \mathrm{I}_d)$ and $p_0$ being $N_d(0, \mathrm{I}_d)$ (for the size study) or $T_{d}(5)$ (for the power study).

\begin{figure}[!ht]
\centering
\includegraphics[width=37pc,height=16pc]{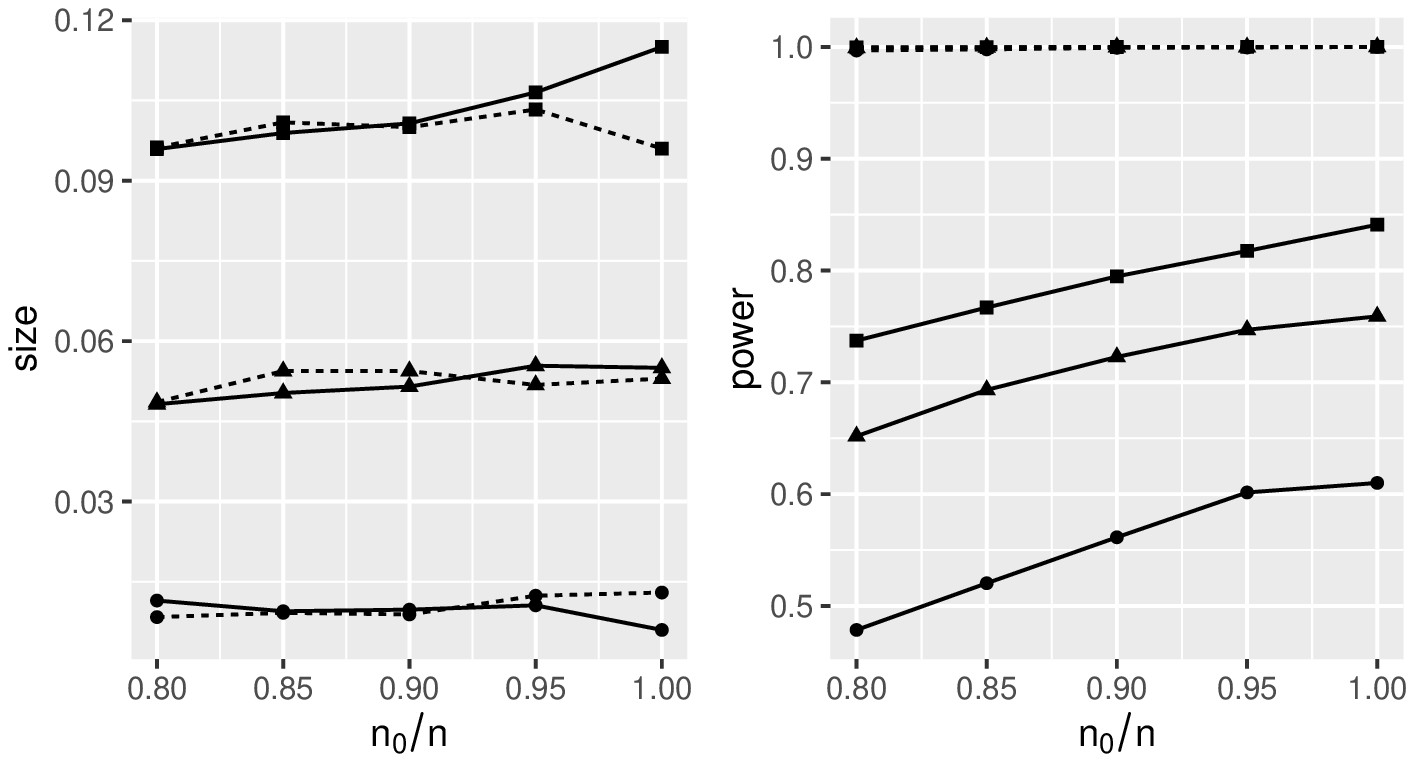}
\includegraphics[width=37pc,height=16pc]{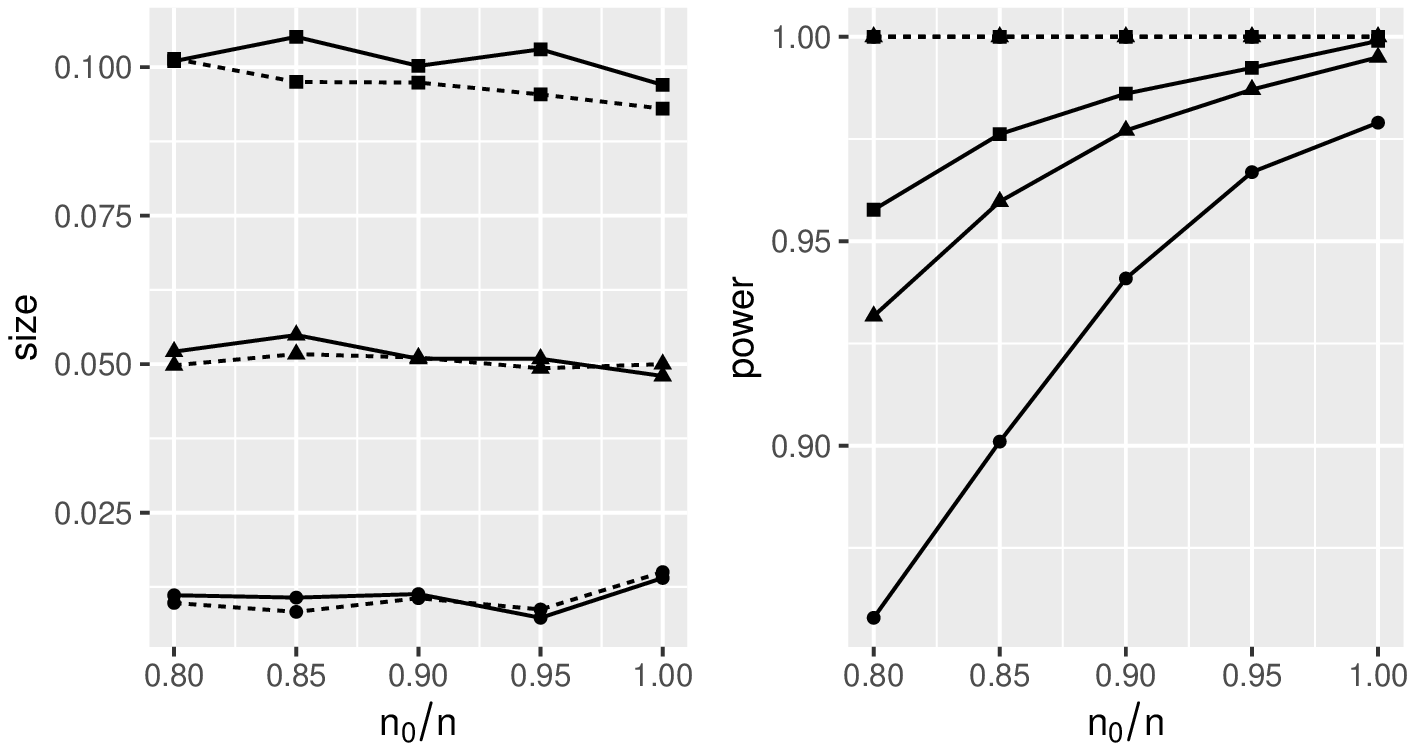}
\caption{The size and power of $\widehat{\mathbb{S}}$ across the subsample ratio $n_0/n$ for $n=100$ (solid line) and $n=500$ (dotted line), where the significance level is
1\% (circle points), 5\% (diamond points), and 10\% (square points). Top panels: $d=2$; Bottom panels: $d=5$.}
\label{sensitive_sigma}
\end{figure}

First, we consider the cases that $n_0$ is taken with the subsample ratio $n_0/n=0.8$, $0.9$, $0.95$, and $1$, while the
value of $\sigma$ is chosen as in (\ref{sigma_choice}). Fig\,\ref{sensitive_sigma} plots the size and power of $\widehat{\mathbb{S}}$ across
the subsample ratio $n_0/n$. From this figure, we can find that (1) $\widehat{\mathbb{S}}$ always has a good size performance; (2) when $n=100$,
the power of $\widehat{\mathbb{S}}$
increases as the value of $n_0/n$ (or $n_0$) increases, and when $n=500$, the power of $\widehat{\mathbb{S}}$ reaches one in all examined cases.
Therefore, as expected, we should recommend to use $n_0=n$ for $\widehat{\mathbb{S}}$,
although this choice of $n_0$ is inconsistent to our theoretical setting.

\begin{figure}[!ht]
\centering
\includegraphics[width=37pc,height=16pc]{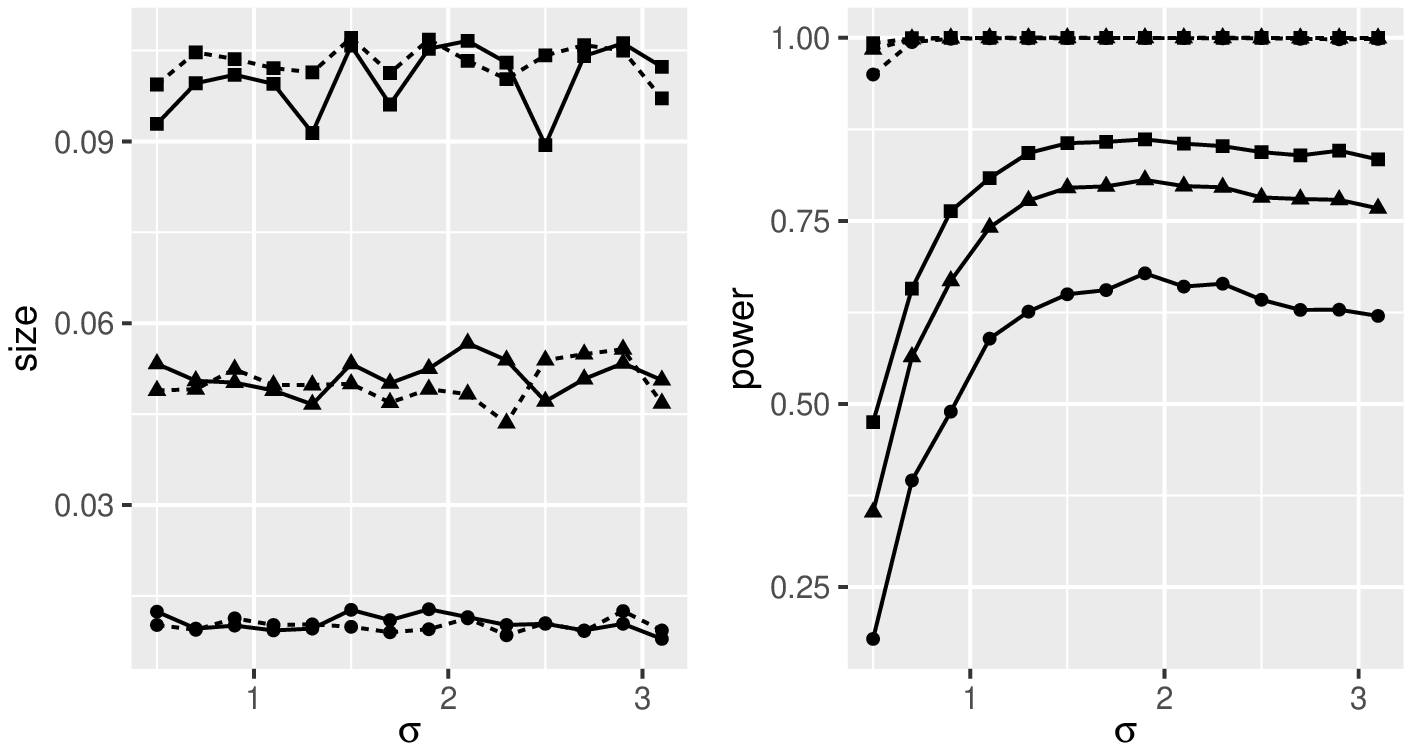}
\includegraphics[width=37pc,height=16pc]{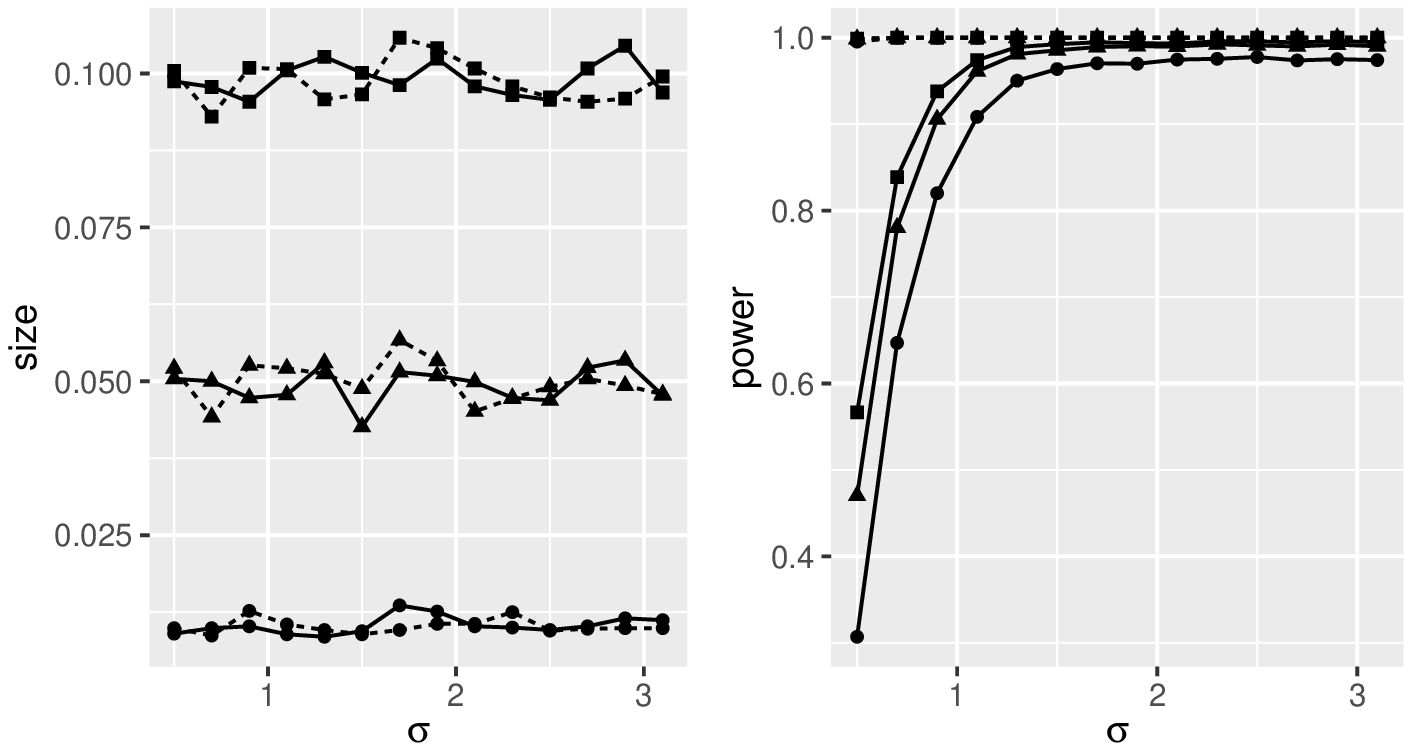}
\caption{The size and power of $\widehat{\mathbb{S}}$ across the kernel parameter $\sigma$ for $n=100$ (solid line) and $n=500$ (dotted line), where the significance level is
1\% (circle points), 5\% (diamond points), and 10\% (square points).
Top panels: $d=2$; Bottom panels: $d=5$.
}
\label{sensitive_n0}
\end{figure}

Second, we consider the cases that $\sigma$ is set to be 0.5, 0.7, ..., 3.1, while the value of $n_0$ is taken as $n$.
Fig\,\ref{sensitive_n0} plots the size and power of $\widehat{\mathbb{S}}$ across $\sigma$. From this figure, we can find that
the size of $\widehat{\mathbb{S}}$ is always accurate for each examined $\sigma$, and the power of $\widehat{\mathbb{S}}$ for $\sigma\geq 1.1$ has only a marginal difference
from that for the choice of $\sigma$ in (\ref{sigma_choice}). These findings imply that $\widehat{\mathbb{S}}$ tends to have a stable size and power performance
over the choice of $\sigma$.

\section{Application}
\begin{figure}[!ht]
\centering
\includegraphics[scale=1]{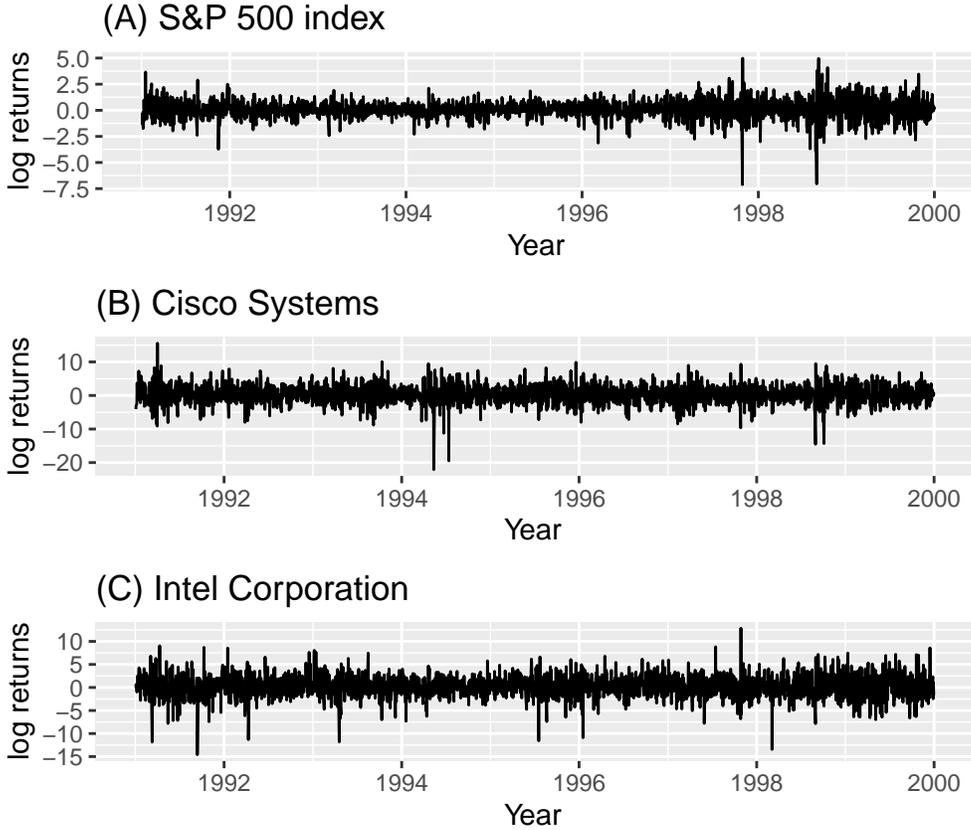}
\caption{The daily log returns (in percentage) of S\&P 500 index, Cisco Systems, and Intel Corporation.}
\label{fig:time plot}
\end{figure}

In this section, we revisit a real example in Tsay (2005). This example considered a three-dimensional financial time series, which consists of the daily log returns (in percentage) of the S\&P 500 index, the stock price of Cisco Systems, and the stock price of Intel Corporation from January 2, 1991 to December 31, 1999, with 2275 observations in total.
We denote this multivariate time series by $Y_t=(Y_{1t},Y_{2t},Y_{3t})^\top$, and plot each entry of $Y_t$ in Fig\,\ref{fig:time plot}.
Following Tsay (2005), $Y_t$ is fitted by a VAR(3)--CCC--GARCH(1, 1) model
\begin{equation}\label{fitted_model_application}
\left\{
\begin{array}{l}
Y_t=M+A_1Y_{t-1}+A_2Y_{t-2}+A_3Y_{t-3}+\varepsilon_{t},\\
\varepsilon_{t}=C_{t}^{1/2}\eta_t,\\
C_{t}=\mbox{diag}\{\sigma_{1,t},\sigma_{2,t},\sigma_{3,t}\} \cdot R\cdot \mbox{diag}\{\sigma_{1,t},\sigma_{2,t},\sigma_{3,t}\}
\end{array}
\right.
\end{equation}
with
\begin{flalign*}
\begin{pmatrix}
    \sigma_{1,t}^2 \\
    \sigma_{2,t}^2 \\
    \sigma_{3,t}^2
  \end{pmatrix}
  =  W +
    B
    \begin{pmatrix}
        \varepsilon_{1,t-1}^2\\
        \varepsilon_{2,t-1}^2\\
        \varepsilon_{3,t-1}^2
    \end{pmatrix} +
    \Gamma
\begin{pmatrix}
    \sigma_{1,t-1}^2 \\
    \sigma_{2,t-1}^2 \\
    \sigma_{3,t-1}^2
  \end{pmatrix}.
\end{flalign*}
For model (\ref{fitted_model_application}), after dropping the insignificant parameters, we follow Tsay (2005) to first estimate the VAR(3) model by using the LSE, and
then estimate the CCC--GARCH(1, 1) model by using the QMLE, where the resulting estimators are given by
\begin{flalign*}
\widehat{M} &= \begin{pmatrix}
    0.071 \\
    0.275\\
    0.164
    \end{pmatrix},\quad\quad\quad\quad\quad\quad\quad\,\,\,\,\,\,\,\,\,
\widehat{A}_1 = \begin{pmatrix}
    0 & 0 & 0\\
    0 & 0 & 0\\
    -0.236 & 0 & 0.053
    \end{pmatrix},\\
\widehat{A}_2 &= \begin{pmatrix}
    0 & 0 & 0\\
    0.282 & -0.122 & 0\\
    0 & 0 & 0
    \end{pmatrix},\quad\,\,\,\,\,\,\,\,\,\,\,\,\,\,
\widehat{A}_3 = \begin{pmatrix}
    -0.054 & 0 & 0\\
    0 & 0 & 0\\
    0 & 0 & 0
    \end{pmatrix},\\
\widehat{R} &= \begin{pmatrix}
    1 & 0.518 & 0.489 \\
    0.518 & 1 & 0.478 \\
    0.489 & 0.478 & 1
\end{pmatrix}, \quad\quad\,\,
\widehat{W}  =   \begin{pmatrix}
        0.004\\
        0.170\\
        0.053
    \end{pmatrix}, \\
\widehat{B} &   =
    \begin{pmatrix}
        0.044 & 0 & 0 \\
        0 & 0.058 & 0.001\\
        0.013 & 0 & 0.017
    \end{pmatrix}, \quad\quad\,\,\,\,
\widehat{\Gamma}
    = \begin{pmatrix}
        0.942 & 0 & 0.001 \\
        0 & 0.921 & 0 \\
        0.001 & 0 & 0.978
    \end{pmatrix}.
\end{flalign*}

Next, we use our KSD-based test $\widehat{\mathbb{S}}$ to check the distribution of $\eta_t$.
The null distribution $p_0(x)$ of interest is $N_{3}(0, \mathrm{I}_3)$, $T_{3}(\nu)$, or $SN_{3}(\gamma)$,
where the degrees of freedom $\nu$ is $\nu_{MLE}$, 6, 7, 8, or 9,  and the skewness vector
$\gamma$ is $\gamma_{MLE}$. Here, $\nu_{MLE}=7.724$ is the maximum likelihood estimator (MLE) of
$\nu$ based on $\eta_t\sim T_{3}(\nu)$, and $\gamma_{MLE}=(-0.181, -0.023, 0)$ is the MLE of $\gamma$ based on $\eta_t\sim SN_{3}(\gamma)$.
To calculate $\widehat{\mathbb{S}}$, we choose $n_0=n$ and use the Gaussian kernel $k$ in (\ref{Gaussian_k}) with $\sigma$ taken as in (\ref{sigma_choice}).
The p-value of $\widehat{\mathbb{S}}$ is computed based on the parametric bootstrap in Subsection 3.3 with  $m=1000$.

Table \ref{table7} reports the p-values of $\widehat{\mathbb{S}}$ for all chosen  null distributions $p_0(x)$. From this table, we
can find that $\widehat{\mathbb{S}}$ gives the strong evidence to reject the null distributions $N_{3}(0, \mathrm{I}_3)$ and $SN_{3}(\gamma_{MLE})$, and
on the contrary, it can not reject the null distributions $T_{3}(\nu_{MLE})$, $T_{3}(7)$, $T_{3}(8)$, and $T_{3}(9)$ at the significant level 5\%.
Since $\widehat{\mathbb{S}}$ has the largest p-value for the null distribution $T_{3}(8)$,
it is reasonable to conclude that $\eta_t$ in model (\ref{fitted_model_application}) follows $T_{3}(8)$.

\begin{table}[!ht]
\caption{The p-values of $\widehat{\mathbb{S}}$ for different null distributions $p_0(x)$}\label{table7}\centering
\addtolength{\tabcolsep}{-0.1pt}
 \renewcommand{\arraystretch}{1.1}
\begin{threeparttable}

\begin{tabular}{@{\extracolsep{\fill}}c|ccccccc}
\hline
$p_0(x)$      &  $N_{3}(0, \mathrm{I}_3)$  & $T_{3}(\nu_{MLE})$  &  $T_{3}(6)$ & $T_{3}(7)$ & $T_{3}(8)$ & $T_{3}(9)$ &  $SN_{3}(\gamma_{MLE})$   \bigstrut[t]\\
\hline
\mbox{p-value} &  0.000  & 0.148 & 0.025  & 0.067 & 0.152 & 0.062 & 0.000 \bigstrut[b]\\
\hline
\end{tabular}%


\end{threeparttable}
\end{table}

\section{Concluding remarks}
This paper constructed a new KSD-based test to detect the error distribution in multivariate time series models with general specifications.
The KSD-based test is easy-to-implement as long as the (Stein) score function of the null distribution has an explicit form.
Hence, it allows the null distribution of interest to be not only multivariate normal, but also multivariate $t_{\nu}$, skew-normal, and many others.
Since most of the existing tests only deal with the multivariate normal null distribution, the KSD-based test can largely
broaden the testing scope for practitioners. This progress driven by the KSD-based test is important in view of the fact
that the non-normal distributed errors are often recommended in various economic and financial applications.

Furthermore, our extensive simulation studies found that
the KSD-based test not only shows its generality advantage to deal with multivariate non-normal null distributions, but also
exhibits the comparative power with the existing tests to handle the multivariate normal null distribution.
Finally, we studied a 3-dimensional financial time series by a VAR(3)--CCC--GARCH(1, 1) model, and the results of KSD-based test
indicated that the error of this model follows a 3-dimensional multivariate $t_{8}$ distribution.

\renewcommand{\thesection}{A}
\setcounter{equation}{0}
\setcounter{section}{0}
\section*{Appendices} 

\subsection{The expansion of $\widehat{\mathbb{S}}$}

To facilitate our proofs,  we need a useful expansion of $\widehat{\mathbb{S}}$.
First, we give some notation to present this expansion.
Denote $\zeta_n=\widehat\theta_n-\theta_0$ and
\begin{flalign*}
&\varsigma_i^{(1)}=\big(\eta_i, \nabla_{\theta} g_i(\theta_0)\big)\in \mathbb{R}^{d\times1}\times \mathbb{R}^{d\times q}, \\
&\varsigma_i^{(2)}=\big(\eta_i, \nabla_{\theta} g_i(\theta_0), \nabla_{\theta} \text{vec}(\nabla_{\theta}g_i(\theta_0)\big)\in \mathbb{R}^{d\times1}\times \mathbb{R}^{d\times q}\times \mathbb{R}^{qd\times q},\\
&G_{ij}(\theta)=(g_{i}(\theta)^\top, g_{j}(\theta)^\top)^\top\in \mathbb{R}^{2d\times 1},\\
&\eta_{ij}=(\eta_i^\top,\eta_j^\top)^\top\in \mathbb{R}^{2d\times 1}, \,\,\,\,\,\,\,\widehat{\eta}_{ij}=(\widehat{\eta}_i^\top,\widehat{\eta}_j^\top)^\top\in \mathbb{R}^{2d\times 1},\\
&W_{ij}=W(\eta_{ij})\in\mathbb{R}^{2d\times 1},\,\,\,\,\,\,\,\,\,\,H_{ij}=H(\eta_{ij})\in\mathbb{R}^{2d\times 2d},
\end{flalign*}
where
\begin{flalign}
W(x,x')&=\left (\nabla_{x} u(x,x')^{\top}, \nabla_{x'} u(x,x')^{\top}\right )^\top\in \mathbb{R}^{2d\times1},\\
H(x,x')&=\left(\begin{matrix}
\nabla_{x,x} u(x,x') & \nabla_{x,x'} u(x,x')\\
\nabla_{x,x'} u(x,x') & \nabla_{x',x'} u(x,x')
\end{matrix}\right)\in \mathbb{R}^{2d\times 2d}.
\end{flalign}

Second, we define three U-statistics $\mathbb{S}^{(a)}$ (for $a=0, 1, 2$) as follows:
\begin{equation}\label{eq:1.1}
\mathbb{S}^{(a)}=\dfrac{2}{n_0(n_0-1)}\sum_{n-n_0+1\le i<j\le n}h^{(a)}(\varsigma_i^{(a)}, \varsigma_j^{(a)}),
\end{equation}
where
\begin{align*}
&h^{(0)}(\varsigma_i^{(0)}, \varsigma_j^{(0)})=u(\eta_i, \eta_j),\,\,\,\,
h^{(1)}(\varsigma_i^{(1)}, \varsigma_j^{(1)})=\big(\nabla_{\theta}G_{ij}(\theta_0)^{\top}\big)W_{ij}
\in \mathbb{R}^{q\times1},\\
&\,\,\,\,\mbox{ and } \,\,\,\,
h^{(2)}(\varsigma_i^{(2)}, \varsigma_j^{(2)})=\big ( b_{rs} \big)_{q\times q}\in \mathbb{R}^{q\times q}
\end{align*}
with $b_{rs}=\big(\nabla_{\theta_r} G_{ij}(\theta_0)^\top\big)H_{ij}\big(\nabla_{\theta_s} G_{ij}(\theta_0)\big).$

With these notation, by Taylor's expansion we have
\begin{equation}\label{eqA1}
\begin{split}
u(\widehat\eta_i,\widehat\eta_j)
&=u(\eta_i,\eta_j)+(\widehat\eta_{ij}-\eta_{ij})^\top W_{ij}+\dfrac{1}{2}(\widehat\eta_{ij}-\eta_{ij})^\top H_{ij}(\widehat\eta_{ij}-\eta_{ij})+R_{ij}^{(1)},
\end{split}
\end{equation}
where  $H_{ij}^{\dagger}=H(\eta_{ij}^\dagger)$, $\eta_{ij}^\dagger$ lies between $\eta_{ij}$ and $\widehat\eta_{ij}$, and
\begin{equation*} 
R_{ij}^{(1)}=\dfrac{1}{2}(\widehat\eta_{ij}-\eta_{ij})^\top(H_{ij}^\dagger-H_{ij})(\widehat\eta_{ij}-\eta_{ij}).
\end{equation*}
Furthermore, by Taylor's expansion again we have
\begin{align} \label{eqA4}
\widehat\eta_{ij}-\eta_{ij}&=\bar{R}_{ij}^{(2)}+\big(\nabla_{\theta}G_{ij}(\theta^{\dagger})\big)\zeta_n
=\bar{R}_{ij}^{(2)}+\bar{R}_{ij}^{(3)}+\big(\nabla_{\theta}G_{ij}(\theta_0)\big)\zeta_n,
\end{align}
where $\theta^\dagger$ lies between $\theta_0$ and $\widehat\theta_n$, and
 \begin{align*} 
 \bar{R}_{ij}^{(2)}=(\widehat{R}_{i}(\widehat\theta_{n})^\top, \widehat{R}_{j}(\widehat\theta_{n})^\top)^\top,\,\,\,\,
 \bar{R}_{ij}^{(3)}=\big(\nabla_{\theta}G_{ij}(\theta^{\dagger})-\nabla_{\theta}G_{ij}(\theta_0)\big)\zeta_n.
 \end{align*}
By (\ref{E4}) and (\ref{eqA1})--(\ref{eqA4}), it follows that
\begin{equation}\label{A-8}
\widehat{\mathbb{S}}=\mathbb{S}^{(0)}+\zeta_n^\top\mathbb{S}^{(1)}+\dfrac{1}{2}\zeta_n^\top\mathbb{S}^{(2)}\zeta_n+\widehat{R},
\end{equation}
where the U-statistics $\mathbb{S}^{(a)}$ (for $a=0, 1, 2$) are defined in (\ref{eq:1.1}), and the remainder term
$\widehat{R}$ is defined by
\begin{equation}\label{eqA7}
\widehat{R}=\dfrac{2}{n_0(n_0-1)}\sum_{n-n_0+1\le i<j\le n} R_{ij}
\end{equation}
with $R_{ij}=R_{ij}^{(1)}+R_{ij}^{(2)}$ and
\begin{equation*} 
R_{ij}^{(2)}=(\bar R_{ij}^{(2)}+\bar R_{ij}^{(3)})^\top W_{ij}+
\Big[\dfrac{1}{2}(\bar R_{ij}^{(2)}+\bar R_{ij}^{(3)})^\top+\zeta_n^\top\big(\nabla_{\theta} G_{ij}(\theta_0)^{\top}\big) \Big] H_{ij}(\bar R_{ij}^{(2)}+\bar R_{ij}^{(3)}).
\end{equation*}

From the expansion (\ref{A-8}), it is clear that the estimation effect has an impact on the limiting distribution of
$\widehat{\mathbb{S}}$ through the linear term $\zeta_n^\top\mathbb{S}^{(1)}$, the quadratic term $\zeta_n^\top\mathbb{S}^{(2)}\zeta_n$, and the remainder term $\widehat{R}$, and that the effect of unobserved initial values is involved in the remainder term $\widehat{R}$ via $\bar{R}_{ij}^{(2)}$.

\subsection{Proofs of Theorems \ref{thm1}--\ref{thm2}}

To prove Theorems \ref{thm1}--\ref{thm2}, we need two technical lemmas to handle the
effects of estimation uncertainty and unobserved initial values.

\begin{lem}\label{lem1}
	Suppose Assumptions \ref{asm1}, \ref{asm3}--\ref{asm4}, \ref{asm6} and \ref{asm7}(ii) hold.
Then, 

	(i) $n_0\zeta_n^\top\mathbb{S}^{(1)}=o_{p}(1)$, provided that $\varepsilon>1/2$;

     (ii) $n_0\zeta_n^\top\mathbb{S}^{(2)}\zeta_n=o_{p}(1)$, provided that $\varepsilon>0$.
\end{lem}


\begin{lem}\label{lem3}
	Suppose Assumptions \ref{asm1}--\ref{asm7} hold. Then,
$$n_0\widehat R=o_p(1), \mbox{ provided that }\varepsilon>0,$$
where $\widehat R$ is defined in (\ref{eqA7}).
\end{lem}

\noindent \textbf{Proof of Lemma \ref{lem1}}. By Assumptions \ref{asm1}, \ref{asm3}, \ref{asm6} and \ref{asm7}(ii) and
the law of large numbers for U-statistics, it is not hard to see that
$\mathbb{S}^{(1)}=O_{p}(1)$ and $\mathbb{S}^{(2)}=O_{p}(1)$.
Since $\sqrt{n}\zeta_n=O_p(1)$ by Assumption \ref{asm4}, it follows that
$n_0\zeta_n^\top\mathbb{S}^{(1)}=O_{p}(n_0/\sqrt{n})=O_{p}(1/n^{\varepsilon-1/2})$
and $n_0\zeta_n^\top\mathbb{S}^{(2)}\zeta_n=O_{p}(n_0/n)=O_{p}(1/n^{\varepsilon})$.
Hence, the conclusions hold. \qed

\vspace{2mm}

\noindent \textbf{Proof of Lemma \ref{lem3}}. For simplicity, we only show that
\begin{equation}\label{key}
\dfrac{2}{n_0-1}\sum_{n-n_0+1\le i<j\le n}R^{(1)}_{ij}=o_{p}(1),
\end{equation}
since the proof for $R^{(2)}_{ij}$ is similar and even simpler. By (\ref{eqA4}), we can rewrite $R^{(1)}_{ij}$ as
\begin{equation*} 
\begin{split}
R^{(1)}_{ij}&=\dfrac{1}{2}[\bar R^{(2)}_{ij}]^\top (H^\dagger_{ij}-H_{ij})\bar R^{(2)}_{ij}+\dfrac{1}{2}[\bar R^{(3)}_{ij}]^\top (H^\dagger_{ij}-H_{ij})\bar R^{(3)}_{ij}\\&\quad+\dfrac{1}{2}\zeta_n^T  (H^\dagger_{ij}-H_{ij})\zeta_n+[\bar R^{(2)}_{ij}]^\top (H^\dagger_{ij}-H_{ij})\bar R^{(3)}_{ij}\\
&\quad+[\bar R^{(2)}_{ij}]^\top (H^\dagger_{ij}-H_{ij})\big(\nabla_{\theta} G_{ij}(\theta_0)\big)\zeta_n+[\bar R^{(3)}_{ij}]^\top (H^\dagger_{ij}-H_{ij})\big(\nabla_{\theta} G_{ij}(\theta_0)\big)\zeta_n\\
&=:r^{(1)}_{1,ij}+r^{(1)}_{2,ij}+r^{(1)}_{3,ij}+r^{(1)}_{4,ij}+r^{(1)}_{5,ij}+r^{(1)}_{6,ij}.
\end{split}
\end{equation*}
Then, it follows that
$2(n_0-1)^{-1}\sum_{n-n_0+1\le i<j\le n}R^{(1)}_{ij}=\sum_{a=1}^6 \Delta_a^{(1)}$,
where
$$\Delta_a^{(1)}=\dfrac{2}{n_0-1}\sum_{n-n_0+1\le i<j\le n} r^{(1)}_{a,ij}.$$

Let $K>0$ be a generic constant whose value may change from place to place.
Next, we show that $\Delta_1^{(1)}=o_{p}(1)$. To facilitate it, we claim
\begin{equation}\label{eqA23}
\norm{H_{ij}^\dagger-H_{ij}}\le K \left(\norm{\bar R_{ij}^{(2)}}+\norm{\eta_{ij}}+1\right)^2+o_p(1),
\end{equation}
where $o_p(1)$ holds uniformly in $i,j$. With loss of generality, we prove (\ref{eqA23}) for
$\nabla_{x,x}u(\hat\eta_i^\dagger,\hat\eta_j^\dagger)-\nabla_{x,x}u(\eta_i,\eta_j)$, the first block
entry of $H_{ij}^\dagger-H_{ij}$. Denote
$u(x,x'):=u_1(x,x')+u_2(x,x')+u_3(x,x')+u_4(x,x')$, where
$u_1(x,x')=s_{p_0}(x)^\top k(x,x')s_{p_0}(x')$, $u_2(x,x')=s_{p_0}(x)^\top k_{x'}(x,x')$, $u_3(x,x')=k_x(x,x')^\top s_{p_0}(x')$, and
$u_4(x,x')=\text{trace}\big(k_{xx'}(x,x')\big)$. Below, we first prove
\begin{equation}\label{eqA20}
\norm{\nabla_{x,x}u_1(\widehat\eta_i^\dagger,\widehat\eta_j^\dagger)-\nabla_{x,x}u_1(\eta_i,\eta_j)}\le K \left(\norm{\bar R_{ij}^{(2)}}+\norm{\eta_{ij}}+1\right)^2+o_p(1).
\end{equation}
Rewrite
\begin{equation*}
\begin{split}
& \nabla_{x,x} u_1(x,x')=\nabla_{x,x} [s_{p_0}(x)^\top k(x,x')s_{p_0}(x')]=\sum_{1\le r,s\le d}\nabla_{x,x} [s_{p_0}^{(r)}(x)k(x,x')s_{p_0}^{(s)}(x')]\\
&=\sum_{1\le r,s\le d}\bigg (\nabla_{x,x} [s_{p_0}^{(r)}(x)]k(x,x')s_{p_0}^{(s)}(x')+\nabla_{x,x} [k(x,x')]s_{p_0}^{(r)}(x)s_{p_0}^{(s)}(x')\\
&\quad+[\nabla_{x} k(x,x')][\nabla_{x}s_{p_0}^{(r)}(x)]^\top s_{p_0}^{(s)}(x')+[\nabla_{x}s_{p_0}^{(r)}(x)][\nabla_{x}k(x,x')]^\top s_{p_0}^{(s)}(x')
\bigg)\\
&=:\sum_{1\le r,s\le d}\left (T_1^{(r,s)}(x,x')+T_2^{(r,s)}(x,x')+T_3^{(r,s)}(x,x')+T_4^{(r,s)}(x,x')\right).
\end{split}
\end{equation*}
Note that
\begin{align}
\|\widehat\eta_t^\dagger-\eta_t\|&\le \|\widehat R_t(\widehat\theta_n)\|+\|\widehat\theta_n-\theta_0\|\sup_{\theta}\|\nabla_{\theta} g_t(\theta)\|\nonumber\\
&=\|\widehat R_t(\widehat\theta_n)\|+o_p(1),\label{aa_1}\\
\norm{f(x)}&<K(\norm{x}+1) \mbox{ for }f=s_{p_0}(x),\nabla_{x}s_{p_0}(x),
\nabla_{x,x}s_{p_0}(x),\label{aa_2}
\end{align}
where $o_p(1)$ in (\ref{aa_1}) holds uniformly in $t$ due to the fact that $\sqrt{n}\|\widehat\theta_n-\theta_0\|=O_p(1)$ and $\nonumber
n^{-1/2}\max_{1\le t\le n}\sup_{\theta}\norm{\nabla_{\theta} g_t(\theta)}=o_p(1)
$
by Assumptions \ref{asm3}(i) and \ref{asm4}, and (\ref{aa_2}) holds by
Assumption  \ref{asm6}. Therefore, by (\ref{aa_1})--(\ref{aa_2}) and Assumption \ref{asm7},
the adding and subtracting arguments give us
\begin{equation}\nonumber
\begin{split}
&\norm{T_1^{(r,s)}(\widehat\eta_i^\dagger,\widehat\eta_j^\dagger)-T_1^{(r,s)}(\eta_i,\eta_j)}\\
=&\norm{\nabla_{x,x}[s_{p_0}^{(r)}(\widehat\eta_i^\dagger)]k(\widehat\eta_i^\dagger,\widehat\eta_j^\dagger)s_{p_0}^{(s)}(\widehat\eta_j^\dagger)
-\nabla_{x,x}[s_{p_0}^{(r)}(\eta_i)]k(\eta_i,\eta_j)s_{p_0}^{(s)}(\eta_j)}\\
\le& K \left(\norm{\bar R_{ij}^{(2)}}+\norm{\eta_{ij}}+1\right)^2+o_p(1).
\end{split}
\end{equation}
Similarly, the same result holds for
$\|T_b^{(r,s)}(\widehat\eta_i^\dagger,\widehat\eta_j^\dagger)-T_b^{(r,s)}(\eta_i,\eta_j)\|$
with $b=2,3,4.$
Hence, the result (\ref{eqA20}) holds, and then we can show the same result for
$u_{b}(\cdot,\cdot)$ with $b=2,3,4.$
Therefore, it entails that the result (\ref{eqA23}) holds.

Note that $E\norm{\eta_{ij}}^4\le K(E\norm{\eta_{i}}^4+E\norm{\eta_{j}}^4)<\infty$ by Assumption \ref{asm2},
$E\|\bar R_{ij}^{(2)}\|^4\le K(E\|\bar R_{i}^{(2)}\|^4+E\{\bar R_{j}^{(2)}\|^4)$, and
\begin{equation}\nonumber
\lim\limits_{n\rightarrow\infty}\sum_{n-n_0+1\le i \le n}E\|\bar R_{ij}^{(2)}\|^4=0\,\,\mbox{ for all }j
\end{equation}
by Assumption \ref{asm5}. Hence, by (\ref{eqA23}) and H\"older's inequality, we can show
\begin{equation} \label{eqA24}
E|\Delta_1^{(1)}|\leq\frac{1}{n_0-1}\sum_{n-n_0+1\le i<j\le n}E\left [\norm{\bar R_{ij}^{(2)}}^2 \left(\norm{\bar R_{ij}^{(2)}}+\norm{\eta_{ij}}+1\right)^2\right ]=o(1),
\end{equation}
implying that $\Delta_1^{(1)}=o_{p}(1)$.

Furthermore,  by Taylor's expansion, Assumptions \ref{asm3}--\ref{asm4}, and a similar argument as for (\ref{aa_1}), it is straightforward to see
\begin{equation}\label{key_1}
\begin{split}
\norm{\bar R_{ij}^{(3)}}&\le\norm{\nabla_{\theta} G_{ij}(\theta^\dagger)-\nabla_{\theta} G_{ij}(\theta_0)}\times \norm{\widehat\theta_n-\theta_0}\\
&\le\left[2 \max_{1\le t\le n}\sup_{\theta}\norm{\nabla_{\theta,\theta} g_t(\theta)}\right]\times \norm{\widehat\theta_n-\theta_0}^2=o_p\Big(\dfrac{1}{\sqrt {n}}\Big),
\end{split}
\end{equation}
where $o_p(1)$ holds uniformly in $i,j$. By (\ref{key_1}) and the similar arguments as for (\ref{eqA24}),
we can show that $\Delta_a^{(1)}=o_p(1)$ for $2\leq a\leq 6$. Therefore, it follows that the result (\ref{key}) holds. This completes the proof. \qed

\vspace{2mm}

\noindent \textbf{Proof of Theorem \ref{thm1}}.
By (\ref{A-8}) and Lemmas \ref{lem1}--\ref{lem3},
$n_0\widehat{\mathbb{S}}=n_0{\mathbb{S}}^{(0)}+o_{p}(1)$,
and the result follows by Theorem 4.1(2) in Liu et al. (2016).
This completes the proof. \qed

\vspace{2mm}

\noindent \textbf{Proof of Theorem \ref{thm2}}.
By (\ref{A-8}) and Lemmas \ref{lem1}(ii) and \ref{lem3},
\begin{equation} \label{eq_18}
\sqrt{n_0}\widehat{\mathbb{S}}=\sqrt{n_0}\big(\widehat{\mathbb{S}}-\mathbb{S}(p,p_0)\big)+
\sqrt{n}\zeta_n^\top\sqrt{\frac{n_0}{n}} \mathbb{S}^{(1)}+\sqrt{n_0}\mathbb{S}(p,p_0)+o_{p}(1).
\end{equation}
Now,  the conclusion holds since $\sqrt{n_0}\big(\widehat{\mathbb{S}}-\mathbb{S}(p,p_0)\big)=O_{p}(1)$ by Theorem 4.1(1) in Liu et al. (2016),
$\sqrt{n}\zeta_n=O_{p}(1)$, $\mathbb{S}^{(1)}=O_{p}(1)$, and $\mathbb{S}(p,p_0)>0$ under $H_1$.
This completes the proof. \qed


\subsection{Tests used in simulation studies}

\textbf{1. Mardia's tests}.
Consider the null hypothesis that
\begin{align}\label{a_16}
H_0: \,Y_t\sim_{i.i.d.} \mbox{multivariate normal }N_{d}(\mu, \Sigma).
\end{align}
Mardia (1974) detected $H_0$ in (\ref{a_16}) by proposing the following two test statistics:
$$\widehat{\mathbb{T}}_{M,1} = \frac{1}{n^2} \sum_{i=1}^n \sum_{j=1}^n m_{ij}^3 \quad
\mbox{and}
\quad \widehat{\mathbb{T}}_{M,2} = \frac{1}{n} \sum_{i=1}^n m_{ii}^2,$$
where $m_{ij} = (Y_i - \bar{Y})^\top S_{Y}^{-1}(Y_j - \bar{Y})$, and $\bar{Y}$ and $S_{Y}$ are the sample mean and variance of
$\{Y_t\}_{t=1}^{n}$, respectively. The tests $\widehat{\mathbb{T}}_{1,M}$ and $\widehat{\mathbb{T}}_{2,M}$ make
use of the multivariate extensions of skewness and kurtosis measures in Mardia (1970), and they have the
following limiting null distributions
$$(n/6)\widehat{\mathbb{T}}_{M,1}\xrightarrow{d} \chi^2_{d(d+1)(d+2)/6}
\mbox{ and }\widehat{\mathbb{T}}_{M,2}\xrightarrow{d} N(d(d+2), 8d(d+2)/n).$$

\textbf{2. Doornik--Hansen test}.
Let $s = m_3/m_2^{3/2}$ and $k = m_4/m_2^2$ be the original sample skewness and kurtosis, where $m_j = \frac{1}{n}\sum_{i=1}^n(Y_i - \bar{Y})^j$.
Next, transform $s$ and $k$ into $z_1$ and $z_2$, respectively, where
$$z_1 = \delta \log(y+\sqrt{y^2-1})\mbox{ and }z_2 =\sqrt{9\alpha}\Big(\frac{1}{9\alpha}-1+\sqrt[3]{\frac{\chi}{2\alpha}}\Big).$$
Here,
\begin{align*}
y &= s\sqrt{\frac{(\omega^2-1)(n+1)(n+3)}{12(n-2)}}\mbox{ and }\delta = \frac{1}{\sqrt{\log(\omega^2)}} \mbox{ with } \\
\omega^2 &= -1 + \sqrt{2(\beta-1)} \mbox{ and } \beta = \frac{3(n^2+27n-70)(n+1)(n+3)}{(n-2)(n+5)(n+7)(n+9)}; \\
\alpha &= a + c \cdot s^2 \mbox{ and } \chi = 2l(k-1-s^2) \mbox{ with }\\
a &= \frac{(n-2)(n+5)(n+7)(n^2+27n-70)}{6(n-3)(n+1)(n^2+15n-4)},\\
c &= \frac{(n-7)(n+5)(n+7)(n^2+2n-5)}{6(n-3)(n+1)(n^2+15n-4)},\\
l &= \frac{(n+5)(n+7)(n^3+37n^2+11n-313)}{12(n-3)(n+1)(n^2+15n-4)}.
\end{align*}
Based on $z_1$ and $z_2$, Doornik and Hansen (2008) proposed the test statistic
$\widehat{\mathbb{T}}_{DH}:= z_1^2 + z_2^2$ to detect $H_0$ in (\ref{a_16}), where
the limiting null distribution of $\widehat{\mathbb{T}}_{DH}$ is $\chi^2_{2}$.

\vspace{2mm}

\textbf{3. Henze--Zirkler test}.
To detect $H_0$ in (\ref{a_16}), Henze and Zirkler (1990) proposed a test statistic given by
$$\widehat{\mathbb{T}}_{HZ}:= \frac{1}{n}\sum_{i=1}^n\sum_{j=1}^n e^{-\frac{\beta^2}{2}D_{ij}} - 2(1+\beta^2)^{-\frac{d}{2}}\sum_{i=1}^n e^{-\frac{\beta^2}{2(1+\beta^2)}D_i} + n(1+2\beta^2)^{-\frac{d}{2}},$$
where  $\beta = \frac{1}{\sqrt{2}}\big[\frac{n(2d+1)}{4}\big]^{\frac{1}{d+4}}, $
$D_{ij}=(Y_i - Y_j)^\top S_{Y}^{-1}(Y_i - Y_j)$ is the squared Mahalanobis distance between $Y_i$ and $Y_j$, and
$D_i= (Y_i - \bar{Y})^\top S_{Y}^{-1}(Y_i-\bar{Y})$ is the squared distance of $Y_i$ to the centroid.

Under $H_0$ in (\ref{a_16}), the limiting null distribution of $\widehat{\mathbb{T}}_{HZ}$ is
log-normal with mean $\mu_{HZ}$ and variance $\sigma_{HZ}^2$, where
\begin{align*}
\mu_{HZ} &= 1 - (1+2\beta^2)^{-d/2}\left( 1 + \frac{d\beta^2}{1+2\beta^2} + \frac{d(d+2)\beta^4}{2(1+2\beta^2)^2} \right),\\
\sigma_{HZ}^2 &= 2(1+4\beta^2)^{-d/2}+ 2(1+2\beta^2)^{-d} \left( 1 + \frac{2d\beta^4}{(1+2\beta^2)^2} + \frac{3d(d+2)\beta^8}{4(1+2\beta^2)^4} \right) \\
&\quad - 4 \omega_\beta^{-d/2}\left( 1 + \frac{3d\beta^4}{2\omega_\beta} + \frac{d(d+2)\beta^8}{2\omega_\beta^2} \right)
\end{align*}
with $\omega_\beta = (1+\beta^2)(1 + 3\beta^2)$.
Note that Henze and Zirkler (1990) suggested that this test is proper for sample size $n \geq 20$.

\vspace{2mm}

\textbf{4.  Bai--Chen test}.  For model (\ref{E1}), Bai and Chen (2008) tested the multivariate normal and $t_{\nu}$ distributions for $\eta_t$ by using the martingale transformation. Their testing method requires the explicit formula of $P_0(Y_{it}|Y_{1t,...,Y_{i-1,t}})$ for $i=1,...,d$, where $P_0$ is the c.d.f. of $\eta_t$ under $H_0$ in (\ref{Hypothesis}). However, it is difficult to derive the explicit formula of $P_0(Y_{it}|Y_{1t,...,Y_{i-1,t}})$ for $d>2$, even when $P_0$ is the c.d.f. of multivariate normal or $t_{\nu}$. Below, we only consider the case of $d=2$ as in Bai and Chen (2008).

Partition
 $$M(I_{t-1};\theta) = \left[ \begin{array}{c}
    \mu_{1}(I_{t-1};\theta)\\ \mu_{2}(I_{t-1};\theta)\\
    \end{array} \right ] \mbox{ and }
    C(I_{t-1};\theta) = \left[ \begin{array}{cc}
 \sigma_{1}^2(I_{t-1};\theta) & \sigma_{12}(I_{t-1};\theta)\\
    \sigma_{21}(I_{t-1};\theta) & \sigma_{2}^2(I_{t-1};\theta)\\
    \end{array} \right ].$$
Denote $\widehat{\mu}_{it}=\mu_{i}(\widehat{I}_{t-1};\widehat{\theta}_{n})$, $\widehat{\sigma}_{it}=\sigma_{i}(\widehat{I}_{t-1};\widehat{\theta}_n)$, and
$\widehat{\sigma}_{ij,t}=\sigma_{ij}(\widehat{I}_{t-1};\widehat{\theta}_n)$. Define
\begin{align*}
\widehat{\mathbb{T}}_{BC,1}&=\max\left\{ \sup_r |\widehat{W}_{J,1}(r)|,\sup_r |\widehat{W}_{J,2}(r)| \right\},\\
\widehat{\mathbb{T}}_{BC,2}&= \sup_r |\widehat{W}_{J,1}(r)|+\sup_r |\widehat{W}_{J,2}(r)|,\\
\widehat{\mathbb{T}}_{BC,3}&= \sup_{r} |\widehat{W}_{J,3}(r)|,
\end{align*}
where
\begin{align*}
\widehat{W}_{J,k}(r) &= \widehat{J}_{n,k}(r) - \int_0^r \left [ \dot{g}_k(s)^\top C_k^{-1}(s)\int_s^1\dot{g}_k(\tau)d\widehat{J}_{n,k}(\tau) \right ]ds,
\quad k = 1,2,3,
\end{align*}
with $C_k(s) = \int_s^1 \dot{g}_{k}(r)\dot{g}_{k}^\top(r)dr$, $\dot{g}_{k}(r)$ is the first derivative of $g_{k}(r)$, and
$$\widehat{J}_{n,k}(r) = \frac{1}{\sqrt{n}} \sum_{t=1}^n[I(\widehat{U}_{kt} \leq r) - r] \mbox{ for }k = 1,2,\,\,
\widehat{J}_{n,3}(r)=\frac{1}{\sqrt{2}}\Big[\widehat{J}_{n,1}(r)+\widehat{J}_{n,2}(r)\Big].$$
The choices of $\widehat{U}_{kt}$ and $g_{k}(r)$ are given as follows:

\begin{itemize}
  \item
For testing bivariate normal distribution, we take
\begin{align*}
\widehat{U}_{1t}&= \Phi \left( \frac{Y_{1t} - \widehat{\mu}_{1t}}{\widehat{\sigma}_{1t}} \right),\quad\quad\quad \quad \widehat{U}_{2t} = \Phi \left( \frac{Y_{2t} - \widehat{\mu}_{2|1,t}}{\widehat{\sigma}_{2|1,t}} \right),\\
g_{k}(r) &= (r, \phi(\Phi^{-1}(r)), \phi(\Phi^{-1}(r))\Phi^{-1}(r))^\top\mbox{ for }k=1,2,3,
\end{align*}
where $\widehat{\mu}_{2|1,t}=\widehat{\mu}_{2t}+\widehat{\sigma}_{21,t}\widehat{\sigma}_{1t}^{-2}(Y_{1t}-\widehat{\mu}_{1t})$
and $\widehat{\sigma}_{2|1,t}^2=\widehat{\sigma}_{2t}^2-\widehat{\sigma}_{12,t}^2\widehat{\sigma}_{1t}^{-2}$.

\item
For testing bivariate $t_{\nu}$ distribution, we take
\begin{align*}
\widehat{U}_{1t}&= Q_{\nu} \left( \frac{Y_{1t} - \widehat{\mu}_{1t}}{\sqrt{a_{1\nu}}\widehat{\sigma}_{1t}} \right),\quad\quad\quad \quad \widehat{U}_{2t} = Q_{\nu+1} \left( \frac{Y_{2t} - \widehat{\mu}_{2|1,t}}{\sqrt{a_{2\nu}}\widehat{\sigma}_{2|1,t}} \right),\\
g_1(r) &= (r, q_{\nu}(Q_{\nu}^{-1}(r)), q_{\nu}(Q_{\nu}^{-1}(r))Q_{\nu}^{-1}(r))^\top, \\
g_2(r) &= (r, q_{\nu+1}(Q_{\nu+1}^{-1}(r)), q_{\nu+1}(Q_{\nu+1}^{-1}(r))Q_{\nu+1}^{-1}(r))^\top, \\
g_3(r) &= (r, q_{\nu}(Q_{\nu}^{-1}(r)), q_{\nu}(Q_{\nu}^{-1}(r))Q_{\nu}^{-1}(r), q_{\nu+1}(Q_{\nu+1}^{-1}(r)), q_{\nu+1}(Q_{\nu+1}^{-1}(r))Q_{\nu+1}^{-1}(r))^\top,
\end{align*}
where $Q_{\nu}(x)$ (or $q_{\nu}(x)$) is the c.d.f. (or p.d.f.) of standardized univariate $t_{\nu}$ distribution, and
$$a_{1\nu}=\frac{\nu-2}{\nu},\quad\quad\quad a_{2\nu}=\frac{\nu-2+(Y_{1t} - \widehat{\mu}_{1t})^2\widehat{\sigma}_{1t}^{-2}}{\nu+1}.$$
\end{itemize}

\noindent Note that
$\widehat{\mathbb{T}}_{BC,i}$, $i=1,2,3$, can be computed by using a similar numerical method as in Appendix B of Bai (2003).
Under $H_0$ in (\ref{Hypothesis}), the limiting distributions of $\widehat{\mathbb{T}}_{BC,i}$ can be found in Corollary 3.2 of Bai and Chen (2008).
Let $cv_{BC,i}=(cv_{i,0.01},cv_{i,0.05},cv_{i,0.1})$ be a vector containing the critical values of  $\widehat{\mathbb{T}}_{BC,i}$ at levels
1\%, 5\% and 10\%. By direction simulations, we have that
$cv_{BC,1}=(2.211,2.469,2.993)$,  $cv_{BC,2}=(3.443,3.792,4.504)$, and $cv_{BC,3}=(2.782,2.214,1.940)$.

\vspace{2mm}

\textbf{5. Henze--Jim\'{e}nez-Gamero--Meintanis test}.
When $p_0$ in (\ref{Hypothesis}) is multivariate normal, Henze et al. (2019) made use of the identity (\ref{Henze}) to propose a test statistic given by
\begin{equation*}
    \widehat{\mathbb{T}}_{HJM} =\sqrt{n} \left( \frac{\pi}{\gamma_0}\right)^{d/2} \left( \frac{1}{n^2} \sum_{j,k=1}^n \exp\left(\frac{\rVert \widehat{\eta}_j \rVert^2 - \rVert \widehat{\eta}_k \rVert^2}{4\gamma_0} \right)\cos\left( \frac{{\widehat{\eta}_j}^\top \widehat{\eta}_k}{2\gamma_0}\right) -1 \right),
\end{equation*}
where $\gamma_0>0$ is a fixed constant. As the simulation studies in Henze et al. (2019), we take $\gamma_0=1.5$ and use a similar parametric bootstrap as ours in Subsection 3.3 to compute the
critical values of $\widehat{\mathbb{T}}_{HJM}$.


\begin{center}
{\large \textbf{REFERENCES}}
\end{center}

\indent\hbox to -29.0pt{\hfil} \hangindent=20.0pt \hangafter=1\hskip 10pt
Arellano-Valle, R. B. and Azzalini, A. (2008). The centred parametrization for the multivariate skew-normal distribution.
{\it Journal of Multivariate Analysis} {\bf 99}, 1362--1382.

\indent\hbox to -29.0pt{\hfil} \hangindent=20.0pt \hangafter=1\hskip 10pt
Bai, J. (2003). Testing parametric conditional distributions of dynamic models. {\it Review of Economics and Statistics} {\bf 85}, 531--549.

\indent\hbox to -29.0pt{\hfil} \hangindent=20.0pt \hangafter=1\hskip 10pt
Bai, J. and Chen, Z. (2008). Testing multivariate distributions in GARCH models. {\it Journal of Econometrics} {\bf 143}, 19--36.

\indent\hbox to -29.0pt{\hfil} \hangindent=20.0pt \hangafter=1\hskip 10pt
Bai, J. and Ng, S. (2005). Tests for skewness, kurtosis, and normality for time series data. {\it Journal of Business \& Economic Statistics} {\bf 23}, 49--60.

\indent\hbox to -29.0pt{\hfil} \hangindent=20.0pt \hangafter=1\hskip 10pt
Bauwens, L. and Laurent, S. (2005). A new class of multivariate skew densities, with application to generalized autoregressive conditional heteroscedasticity models. {\it Journal of Business \& Economic Statistics} {\bf 23}, 346--354.

\indent\hbox to -29.0pt{\hfil} \hangindent=20.0pt \hangafter=1\hskip 10pt
Bauwens, L., Laurent, S. and Rombouts, J. V. K. (2006). Multivariate GARCH models: a
survey. {\it Journal of Applied Econometrics} {\bf 21}, 79--109.

\indent\hbox to -29.0pt{\hfil} \hangindent=20.0pt \hangafter=1\hskip 10pt
Berk, J. (1997). Necessary conditions for the CAPM. {\it Journal of Economic Theory} {\bf 73}, 245--257.

\indent\hbox to -29.0pt{\hfil} \hangindent=20.0pt \hangafter=1\hskip 10pt
Bontemps, C. and Meddahi, N. (2005). Testing normality: a GMM approach. {\it Journal of Econometrics} {\bf 124}, 149--186.

\indent\hbox to -29.0pt{\hfil} \hangindent=20.0pt \hangafter=1\hskip 10pt
Bontemps, C. and Meddahi, N. (2012). Testing distributional assumptions: A GMM approach. {\it Journal of Applied Econometrics} {\bf 27}, 978--1012.

%

\indent\hbox to -29.0pt{\hfil} \hangindent=20.0pt \hangafter=1\hskip 10pt
Christoffersen, P. F. and Diebold, F. X. (1997). Optimal prediction under asymmetric loss. {\it Econometric Theory} {\bf 13}, 808--817.

\indent\hbox to -29.0pt{\hfil} \hangindent=20.0pt \hangafter=1\hskip 10pt
Comte, F. and Lieberman, O. (2003). Asymptotic theory for multivariate GARCH processes.
{\it Journal of Multivariate Analysis} {\bf 84}, 61--84.


\indent\hbox to -29.0pt{\hfil} \hangindent=20.0pt \hangafter=1\hskip 10pt
De Luca, G., Genton, M. G. and Loperfido, N. (2006). A multivariate skew-garch model. {\it Advances in Econometrics} {\bf 20}, 33--57.

\indent\hbox to -29.0pt{\hfil} \hangindent=20.0pt \hangafter=1\hskip 10pt
Diebold, F. X., Gunther, T. A. and Tay, A. S. (1998).
Evaluating density forecasts with applications to financial risk management. {\it International Economic Review} {\bf 39}, 863--883.

\indent\hbox to -29.0pt{\hfil} \hangindent=20.0pt \hangafter=1\hskip 10pt
Doornik, J. A. and Hansen, H. (2008). An omnibus test for univariate and multivariate normality. {\it Oxford Bulletin of Economics and Statistics} {\bf 70}, 927--939.

\indent\hbox to -29.0pt{\hfil} \hangindent=20.0pt \hangafter=1\hskip 10pt
Escanciano, J. C. (2006). Goodness-of-fit tests for linear and non-linear time series
models. {\it Journal of the American Statistical Association} {\bf 101}, 531--541.

%
%
%

\indent\hbox to -29.0pt{\hfil} \hangindent=20.0pt \hangafter=1\hskip 10pt
Francq, C., Jim\'{e}nez-Gamero, M. D. and Meintanis, S. G. (2017). Tests for conditional ellipticity in multivariate GARCH models. {\it Journal of Econometrics} {\bf 196}, 305--319.

\indent\hbox to -29.0pt{\hfil} \hangindent=20.0pt \hangafter=1\hskip 10pt
Francq, C. and Zako\"{i}an, J.-M.  (2012). QML estimation of a class of multivariate asymmetric GARCH
models. {\it Econometric Theory} {\bf 28}, 179--206.

\indent\hbox to -29.0pt{\hfil} \hangindent=20.0pt \hangafter=1\hskip 10pt
Francq, C. and Zako\"{i}an, J.-M.   (2019). {\it GARCH Models: Structure, Statistical Inference and
Financial Applications (2nd Edition)}. Wiley, Chichester, UK.


\indent\hbox to -29.0pt{\hfil} \hangindent=20.0pt \hangafter=1\hskip 10pt
Giacomini, R., Politis, D. N. and White, H. (2013). A warp-speed method for conducting
monte carlo experiments involving bootstrap estimators. {\it Econometric Theory} {\bf 29}, 567--589.

\indent\hbox to -29.0pt{\hfil} \hangindent=20.0pt \hangafter=1\hskip 10pt
Haas, M., Mittnik, S. and Paolella, M. S. (2004). Mixed normal conditional heteroskedasticity.
{\it Journal of Financial Econometrics} {\bf 2}, 211--250.

\indent\hbox to -29.0pt{\hfil} \hangindent=20.0pt \hangafter=1\hskip 10pt
Hafner, C. M. and Preminger, A. (2009). On asymptotic theory for multivariate GARCH
models. {\it Journal of Multivariate Analysis} {\bf 100}, 2044--2054.

\indent\hbox to -29.0pt{\hfil} \hangindent=20.0pt \hangafter=1\hskip 10pt
Henze, N., Hl\'{a}vka, Z. and Meintanis, S. G. (2014). Testing for spherical symmetry via the
empirical characteristic function. {\it Statisics} {\bf 48}, 1282--1296.

\indent\hbox to -29.0pt{\hfil} \hangindent=20.0pt \hangafter=1\hskip 10pt
Henze, N., Jim\'{e}nez-Gamero, M. D. and Meintanis, S. G. (2019). Characterizations of multinormality and corresponding tests of fit, including for GARCH models. {\it Econometric Theory} {\bf 35}, 510--546.

\indent\hbox to -29.0pt{\hfil} \hangindent=20.0pt \hangafter=1\hskip 10pt
Henze, N. and Zirkler, B. (1990). A class of invariant consistent tests for multivariate normality.  {\it Communications in Statistics--Theory and Methods} {\bf 19}, 3595--3618.

\indent\hbox to -29.0pt{\hfil} \hangindent=20.0pt \hangafter=1\hskip 10pt
Hong, Y. and Lee, Y. J. (2005). Generalized spectral tests for conditional mean
models in time series with conditional heteroskedasticity of unknown form. {\it Review
of Economic Studies} {\bf 72}, 499--541.

\indent\hbox to -29.0pt{\hfil} \hangindent=20.0pt \hangafter=1\hskip 10pt
Horv\'{a}th, L. and Zitikis, R. (2006). Testing goodness of fit based on densities of GARCH innovations. {\it Econometric Theory} {\bf 22}, 457--482


\indent\hbox to -29.0pt{\hfil} \hangindent=20.0pt \hangafter=1\hskip 10pt
Khmaladze, E. V. (1982). Martingale approach in the theory of goodness-of-fit tests. {\it Theory of Probability \& Its Applications} {\bf 26}, 240--257.

\indent\hbox to -29.0pt{\hfil} \hangindent=20.0pt \hangafter=1\hskip 10pt
Klar, B., Lindner, F. and Meintanis, S. G. (2012). Specification tests for the error distribution in GARCH models. {\it Computational Statistics \& Data Analysis} {\bf 56}, 3587--3598.

\indent\hbox to -29.0pt{\hfil} \hangindent=20.0pt \hangafter=1\hskip 10pt
Koul, H. and Ling, S. (2006). Fitting an error distribution in some heteroscedastic time series models. {\it Annals of Statistics} {\bf 34}, 994--1012.

\indent\hbox to -29.0pt{\hfil} \hangindent=20.0pt \hangafter=1\hskip 10pt
Ling, S. and McAleer, M. (2003). Asymptotic theory for a new vector ARMA-GARCH model.
{\it Econometric Theory} {\bf 19}, 280--310.

\indent\hbox to -29.0pt{\hfil} \hangindent=20.0pt \hangafter=1\hskip 10pt
Liu, Q., Lee, J. and Jordan, M. (2016). A kernelized Stein discrepancy for goodness-of-fit tests. In {\it International Conference on Machine Learning} (pp. 276--284).

\indent\hbox to -29.0pt{\hfil} \hangindent=20.0pt \hangafter=1\hskip 10pt
L\"{u}tkepohl, H. (2005). {\it New Introduction to Multiple Time Series Analysis}. Springer, Berlin.

\indent\hbox to -29.0pt{\hfil} \hangindent=20.0pt \hangafter=1\hskip 10pt
Lobato, I. N. and Velasco, C. (2004). A simple test of normality for time series. {\it Econometric Theory} {\bf 20}, 671--689.


\indent\hbox to -29.0pt{\hfil} \hangindent=20.0pt \hangafter=1\hskip 10pt
Mardia, K. V. (1970). Measures of multivariate skewness and kurtosis with applications.  {\it Biometrika} {\bf 57}, 519--530.

\indent\hbox to -29.0pt{\hfil} \hangindent=20.0pt \hangafter=1\hskip 10pt
Mardia, K. V. (1974). Applications of some measures of multivariate skewness and kurtosis for testing normality and robustness studies.  {\it Sankhy A} {\bf 36}, 115--128.


\indent\hbox to -29.0pt{\hfil} \hangindent=20.0pt \hangafter=1\hskip 10pt
Mecklin, C. J. and Mundfrom, D. J. (2004). An appraisal and bibliography of tests for multivariate normality. {\it International Statistical Review} {\bf 72}, 123--138.

%
%


\indent\hbox to -29.0pt{\hfil} \hangindent=20.0pt \hangafter=1\hskip 10pt
Sz\'{e}kely, G. J. and Rizzo, M. L. (2005). A new test for multivariate normality. {\it Journal of Multivariate Analysis} {\bf 93}, 58--80.

\indent\hbox to -29.0pt{\hfil} \hangindent=20.0pt \hangafter=1\hskip 10pt
Taylor, J. W. (2019). Forecasting value at risk and expected shortfall using a semiparametric approach based on the asymmetric Laplace distribution. {\it Journal of Business \& Economic Statistics} {\bf 37}, 121--133.

\indent\hbox to -29.0pt{\hfil} \hangindent=20.0pt \hangafter=1\hskip 10pt
Tsay, R. S. (2005). {\it Analysis of Financial Time Series}. John Wiley \& Sons, Hoboken, NJ.

\indent\hbox to -29.0pt{\hfil} \hangindent=20.0pt \hangafter=1\hskip 10pt
Tsay, R. S. (2013). {\it Multivariate Time Series Analysis: With R and Financial Applications}. John Wiley \& Sons, Hoboken, NJ.



\indent\hbox to -29.0pt{\hfil} \hangindent=20.0pt \hangafter=1\hskip 10pt
Zhu, K. and Li, W. K. (2015). A new Pearson-type QMLE for conditionally heteroskedastic models. {\it Journal of Business \& Economic Statistics} {\bf 33}, 552--565.

\indent\hbox to -29.0pt{\hfil} \hangindent=20.0pt \hangafter=1\hskip 10pt
Zhu, K. and Ling, S. (2015). Model-based pricing for financial derivatives. {\it Journal of Econometrics} {\bf 187}, 447--457.

\end{document}